\documentclass[11pt]{article}
\usepackage{url}
\usepackage{amsmath,amssymb,amsthm,fullpage,hyperref}
\usepackage{color}
\usepackage{amsfonts}
\usepackage{algpseudocode}
\usepackage[linesnumbered,ruled,vlined]{algorithm2e}
\SetKwProg{Procedure}{Procedure}{}{}
\allowdisplaybreaks

\newtheorem{theorem}{Theorem}[section]
\newtheorem{lemma}[theorem]{Lemma}

\newtheorem{corollary}[theorem]{Corollary}

\newtheorem{prb}{Problem}

\usepackage{comment}

\title{Average Sensitivity of Dynamic Programming} 

 \author{Soh Kumabe\\
 The University of Tokyo\\
 JST, PRESTO\\
 \texttt{soh\_kumabe@mist.i.u-tokyo.ac.jp}
 \and
 Yuichi Yoshida\footnote{Supported by JST, PRESTO Grant Number JPMJPR192B.}\\
 National Institute of Informatics\\
 JST, PRESTO\\
 \texttt{yyoshida@nii.ac.jp}}

\newcommand{\E}{\mathop{\mathbb{E}}}
\newcommand{\EM}{\mathrm{EM}}
\newcommand{\TV}{\mathrm{TV}}
\newcommand{\D}{\mathrm{D}}
\newcommand{\DP}{\texttt{DP}}
\newcommand{\Tmp}{\texttt{Tmp}}

\newcommand{\OPT}{\mathrm{opt}}

\usepackage{fancyhdr} 
\fancypagestyle{pg}
{%
\lhead{}
\rhead{}
\cfoot{--\ \thepage\ --}

}

\begin{document}
\pagestyle{pg}

\maketitle

\begin{abstract}
When processing data with uncertainty, it is desirable that the output of the algorithm is stable against small perturbations in the input.
Varma and Yoshida [SODA'21] recently formalized this idea and proposed the notion of average sensitivity of algorithms, which is roughly speaking, the average Hamming distance between solutions for the original input and that obtained by deleting one element from the input, where the average is taken over the deleted element.

In this work, we consider average sensitivity of algorithms for problems that can be solved by dynamic programming.
We first present a $(1-\delta)$-approximation algorithm for finding a maximum weight chain (MWC) in a transitive directed acyclic graph with average sensitivity $O(\delta^{-1}\log^3 n)$, where $n$ is the number of vertices in the graph.
We then show algorithms with small average sensitivity for various dynamic programming problems by reducing them to the MWC problem while preserving average sensitivity, including the longest increasing subsequence problem, the interval scheduling problem, the longest common subsequence problem, the longest palindromic subsequence problem, the knapsack problem with integral weight, and the RNA folding problem.
For the RNA folding problem, our reduction is highly nontrivial because a naive reduction generates an exponentially large graph, which only provides a trivial average sensitivity bound.

\end{abstract}

\thispagestyle{empty}
\setcounter{page}{0}
\newpage


\section{Introduction}

Dynamic programming (DP) is one of the most successful frameworks for solving practical problems.
From the inception of optimization theory, DP has been used to solve problems in various areas, such as string problems~\cite{palindrome_chowdhury,LCS_hirschberg,LCS_myers,RNA_nussinov,LCS_wagner,RNA_zuker}, scheduling problems~\cite{schedule_chen,schedule_kolen,schedule_xu}, and those of  bioinformatics~\cite{schedule_chen,dan1997algorithms,RNA_nussinov,schedule_xu,RNA_zuker}.
For example, the problem of computing the correspondence between the lines in two text files is often formulated as the \emph{longest common subsequence problem}~\cite{LCS_wagner} or the \emph{longest increasing subsequence problem}~\cite{patiencediff}.

In practice, it is desirable to use ``stable'' algorithms wherein the output does not significantly change by a slight change in the input.
For example, considering a situation in which two people (e.g., Alice and Bob) are editing the same article, after Alice edits the source text file, Bob may compute the correspondence between the lines in the old and new files, and then proofread the edited lines only.
However, if the correspondence between the lines significantly changes after Bob edits a few lines, it is very difficult for him to continue the proofreading.

Varma and Yoshida~\cite{Varma2021} recently introduced the notion of \emph{average sensitivity} to formally argue the stability of an algorithm against input change, and they studied the average sensitivity of algorithms for various graph problems.
Here, the average sensitivity of a (randomized) algorithm $\Call{Alg}{}$ on the input set $V$ of size $n$ is defined as
\begin{align}
    \frac{1}{n}\sum_{i\in V}\EM(\Call{Alg}{V}, \Call{Alg}{V\setminus \{i\}}),
    \label{eq:avg-sensitivity-intro}
\end{align}
where $\Call{Alg}{V}$ and $\Call{Alg}{V \setminus \{i\}}$ denote the distributions of the outputs of $\Call{Alg}{}$ on $V$ and $V \setminus \{i\}$, respectively, and $\EM(\cdot,\cdot)$ is the earth mover's distance~\cite{earthmover_rubner} between two distributions defined as
\begin{align}
  \EM(\mathcal{X}_1,\mathcal{X}_2) := \min_{\mathcal{D}} \E_{(X_1,X_2) \sim \mathcal{D}} |X_1 \triangle X_2|,
  \label{eq:earth-mover}
\end{align}
where the minimum is taken over distributions of a pair of sets such that its marginal distributions on the first and the second coordinates are equal to $\mathcal{X}_1$ and $\mathcal{X}_2$, respectively.
Then, we informally say that an algorithm is \emph{stable on average} if it has a small average sensitivity.

Although it is natural to require stable-on-average algorithms for DP, the naive implementation of a DP is rarely stable on average.
This is because DP iteratively solves subproblems building on the solutions of previously solved subproblems; furthermore, if a change in the input causes a change in the solution for some of the subproblems, it will be propagated and amplified and will cause drastic changes to the final output.

\subsection{Our Contributions}
In this work, we design stable-on-average algorithms for various problems that can be solved by DP\@.
To this end, we first design a stable-on-average algorithm for a problem called the maximum weight chain (MWC) problem, and we then reduce various problems to it.
The details are follows.

\subsubsection{Maximum Weight Chain}
A directed acyclic subgraph (DAG) $G=(V,E)$ is called \emph{transitive} if for any three vertices $v_1,v_2,v_3\in V$ with $(v_1,v_2),(v_2,v_3)\in E$, $(v_1,v_3) \in E$ holds.
In the \emph{MWC problem}, we are given a weighted transitive DAG $G=(V,E,w)$, where $w:V \to \mathbb{R}_+$ is a vertex weight function. The goal is to find a chain\footnote{We save the word ``path'' here because we will use it later in the analysis of the ribonucleic acid (RNA) folding problem.} $P$ of vertices that maximizes the total weight, $\sum_{v \in P}w(v)$.

The MWC problem is a typical problem that can be solved by DP, as in Algorithm~\ref{alg:DP}.
Moreover, it serves as a target problem to which we can reduce various DP problems by regarding each vertex in $G$ as a state of the source DP and each edge in $G$ as the dependency between the states corresponding to the endpoints.
Hence, if we have a stable-on-average algorithm for the MWC problem, we can automatically obtain stable-on-average algorithms for various other problems (unless the generated graph $G$ is exponentially large).

When studying the average sensitivity of algorithms for the MWC problem, we use a slight extension of~\eqref{eq:avg-sensitivity-intro} that is convenient to show reductions from other DP problems.
Let $G=(V,E,w)$ be a transitive DAG, and let $S_1,\dots, S_n$ be \emph{antichains} of $G$, that is, for any $i \in \{1,2,\ldots,n\}$, any two vertices $u,v \in S_i$ do not form an edge in $G$.
Now, the \emph{average sensitivity of an algorithm $\Call{Alg}{}$ for the MWC problem on $G$ with respect to $S_1,\ldots,S_n$} is defined by
\begin{align}
    \frac{1}{n}\sum_{i=1}^{n}\EM(\Call{Alg}{V}, \Call{Alg}{V\setminus S_i}).
    \label{eq:avg-sensitivity}
\end{align}
In the context of reducing some source DP to the MWC problem, $n$ represents the number of elements in the instance of the source DP, and $S_i$ corresponds to the set of the states of the source DP that would disappear if an element $i$ is deleted from the instance.
To emphasize this role of $S_i$, we call each $S_i$ \emph{potentially missing}.


\begin{algorithm}[t!]
\caption{Naive DP for the MWC problem}\label{alg:DP}
\Procedure{\emph{\Call{DynamicProgramming}{$G=(V,E,w)$}}}{
    \For{$v\in V$ in their topological order}{
        \If{$v$ has an incoming edge}{
            $u^* \gets \mathrm{argmax}_{u\in V, (u,v)\in E}w(\DP[u])$\;
            $\DP[v] \gets \DP[u^*]\cup \{v\}$\;
        }\Else{
            $\DP[v] \gets \{v\}$\;
        }
    }
    \Return $\DP\left[\mathrm{argmax}_{v\in V}(w(\DP[v]))\right]$\;
}
\end{algorithm}

In this work, we show that there is an approximation algorithm for the MWC problem with a nontrivially small average sensitivity:
\begin{theorem}\label{thm:main}
    For any $\delta>0$, there is a polynomial-time randomized $(1-\delta)$-approximation algorithm for the MWC problem with the following property:
    Let $G=(V,E,w)$ be a transitive DAG and $S_1,\ldots,S_n$ be antichains such that each vertex in $V$ appears at least one and at most $K$ of $S_1,\ldots,S_n$.
    Then, the average sensitivity of the algorithm on $G$ with respect to $S_1,\ldots,S_n$ is $O(K\delta^{-1}\log^3 |V|)$.
\end{theorem}
We note that the linear dependency on $\delta^{-1}$ is necessary. Let $C>2$ be a constant. 
Consider a disjoint union of the transitive closure of a chain with length $n/2$ and an antichain with size $n/2$.
We set the weights of the vertices in the chain and the antichain as $1$ and $\frac{1-C\delta}{2}n$, respectively. 
Then, any (randomized) $(1-\delta)$-approximation algorithm must output a subset of the chain with probability at least $1/2$. 
However, if we delete random $2C\delta n$ vertices without replacement, then any $(1-\delta)$-approximation algorithm must output a vertex in the antichain with probability at least $1/2$ because the weight of the chain becomes $\frac{1-2C\delta}{2}$ in expectation. 
Therefore, $2C\delta n$ times the average sensitivity is $\Omega(n)$ by the composition theorem of average sensitivity~\cite{Varma2021}; hence, the average sensitivity is $\Omega(\delta^{-1})$. 

It is also natural to consider the \emph{worst-case} sensitivity, which is obtained by replacing the average in~\eqref{eq:avg-sensitivity} with the maximum over $i$'s.
However, there is no algorithm for the MWC problem with a reasonable approximation ratio and a small worst-case sensitivity.
To see this, consider a transitive DAG consisting of the transitive closure of a long chain and an isolated vertex with a large weight.
Although the isolated vertex forms the optimal solution in the original graph, we must use many vertices in the chain for the graph obtained by deleting the isolated vertex.

\subsubsection{Applications}

We can obtain stable-on-average algorithms for various DP problems by reducing them to the MWC problem (with constant $K$ for measuring the average sensitivity).
Here, we describe some representative examples.

\paragraph{Computing Differences Between Two Text Files.}



As discussed, when computing the differences between two text files, it is desirable to use stable-on-average algorithms.
One of the most basic methods is to use the longest common subsequence.
Another popular algorithm is the \emph{patience diff}~\cite{patiencediff}, which focuses more on lines whose copies appear only a small number of times in the files.
This algorithm solves the longest increasing subsequence problem (see Section~\ref{sec:LIS} for the definition) as a subroutine. 

Because both of the longest common subsequence problem and the longest increasing subsequence problem can be solved by textbook DPs that can be formulated as an MWC, Theorem~\ref{thm:main} immediately implies $(1-\delta)$-approximation algorithms with average sensitivity $O(\delta^{-1}\log^3 n)$ for these problems.
See Sections~\ref{sec:LIS} and~\ref{sec:LCS} for more details.

\paragraph{Scheduling and Resource Allocation.}

The \emph{interval scheduling problem} is one the most popular problems for scheduling tasks.
In practice, tasks may be cancelled owing to several reasons such as the lack of participants or a bad weather. 
However, we do not want to drastically change the task schedule because that would incur huge cost.
Hence, it is preferred to have a stable-on-average algorithm for the interval scheduling problem.
Because this problem can be solved by a textbook DP that can be formulated as an MWC\@, Theorem~\ref{thm:main} immediately implies $(1-\delta)$-approximation algorithm with average sensitivity $O(\delta^{-1}\log^3 n)$ for this problem.
See Section~\ref{sec:intervalscheduling} for more details.

Another popular problem used to schedule tasks is the \emph{knapsack problem}.
If the costs of each task and the budget constraint, $C$, are integers, a textbook DP solves the knapsack problem in pseudo-polynomial time, where the pseudo-polynomial factor depends on $C$. 
Because this DP can be formulated as an MWC, Theorem~\ref{thm:main} immediately implies a $(1-\delta)$-approximation algorithm with average sensitivity $O(\delta^{-1}\log^3 (nC))$ for this problem.
See Section~\ref{sec:knapsack} for more details.

\paragraph{Bioinformatics.}
For computational problems in bioinformatics, it is natural to assume that the input has some discrepancy with the true data because the process of observing a biological object injects errors.
Therefore, if we want to obtain a useful output, then the algorithm used must be stable against errors; otherwise the output may not be significantly close to that of the true data. 
One approach to resolving this issue is to design algorithms with small average sensitivity.

Many problems of bioinformatics are solved by DP\@. For example, the interval scheduling problem is applied to the problem of determining protein structure from nuclear magnetic resonance (NMR) peak data~\cite{schedule_chen, schedule_xu}. 
As another example, we note that DP have been used to determine the \emph{secondary structure} of RNA, which represents how the RNA is physically folded.
One popular formulation is the \emph{RNA folding problem} proposed by Nussinov and Jacobson~\cite{RNA_nussinov}.

Besides its usefulness in bioinformatics, the RNA folding problem is theoretically interesting because it is a slight variant of  \emph{two- and one-dimensional (2D/1D) DP}~\cite{galil1992dynamic}, wherein a value of the DP array is given by the maximum over the sum of two values in the DP array.
For example,
\begin{align*}
    \DP[i][j]=\max_{i\leq k < j}(\DP[i][k]+\DP[k+1][j]),
\end{align*}
which cannot be directly formulated as an MWC\@.
To resolve this issue, we introduce a novel technique to formulate such a DP as an MWC using a quasi-polynomial number of vertices, and we show that there is a $(1-\delta)$-approximation algorithm with average sensitivity $O(\delta^{-1}\log^7 n)$.
See Section~\ref{subsec:overview} for a more detailed technical overview and Section~\ref{sec:RNA} for the details of the algorithm and the analysis.

We note that the \emph{longest palindromic subsequence problem} is a special case of the RNA folding problem.
As opposed to the general RNA folding problem, the textbook DP algorithm for this can be directly formulated as an MWC, and hence Theorem~\ref{thm:main} implies a $(1-\delta)$-approximation algorithm with average sensitivity $O(\delta^{-1}\log^3 n)$ for this problem.
See Sections~\ref{sec:LPS} for more details.

\subsection{Related Work}

As mentioned, the notion of average sensitivity was recently proposed by Varma and Yoshida~\cite{Varma2021}.
They studied various graph problems, including the minimum spanning tree problem, minimum cut problem, and maximum matching problem and analyzed the average sensitivities of existing problems as well as developed new algorithms with small average sensitivities.
Zhou and Yoshida~\cite{Yoshida2021} demonstrated a $(1-\epsilon)$-approximation algorithm for the maximum matching problem with sensitivity solely depending on $\epsilon$, where sensitivity is defined as~\eqref{eq:avg-sensitivity} with the average being replaced with the maximum over $i$.
Peng and Yoshida analyzed the average sensitivity of spectral clustering~\cite{Peng2020}, a popular method for graph clustering, and showed that it is stable-on-average if there is a relevant cluster structure in the input graph.

Kiirala~\emph{et~al.}~\cite{kiirala2019safe} investigated stable algorithms for the RNA folding problem. 
Their idea is to enumerate all bases that are paired in all optimal solutions and those that are never paired in any of the optimal solutions, then they construct the output using the enumerated bases.

%

\subsection{Technical Overview}\label{subsec:overview}

\paragraph{Maximum Weight Chain.}
Our algorithm is based on the divide-and-conquer method.
For a vertex $v \in V$ in the input graph $G=(V,E,w)$, let $V_{-v}$ (resp., $V_{+v}$) be the set of vertices $v'$ such that there is a chain from $v'$ to $v$ (resp., from $v$ to $v'$).
Then, we pick up a ``middle'' vertex $\bar{v}$ as a pivot and recurse on the two subgraphs induced by $V_{-\bar{v}}$ and $V_{+\bar{v}}$, respectively.
The output of our algorithm is obtained by concatenating that for $V_{-\bar{v}}$, the vertex $\bar{v}$, and that for $V_{+\bar{v}}$ in this order.

We use the exponential mechanism~\cite{mcsherry2007mechanism} to select the pivot $\bar{v}$.
Specifically, we sample a vertex $v \in V$ as a pivot with probability proportional to $\exp(r(v)/c)$, where $r(v)$ is the maximum weight of a chain containing $v$, and $c$ is an appropriately chosen constant.

We first discuss the approximation ratio.
Because the maximum $r(v)$ over $v\in V$ equals the optimal value of the original instance, in expectation, the optimal value does not decrease much by forcing $v$ to be in the output.
Indeed, for some appropriate choice of $c$, we can prove that the expected value of $r(v)$ is at least $1-\epsilon$ times the optimal value of the original instance, where $\epsilon=O(\frac{\delta}{\log |V|})$
This means that, intuitively, one depth deeper in the recursion decreases the approximation ratio by $\epsilon$.
Hence, if the depth of recursion were to be $O(\log |V|)$, then the approximation ratio would be bounded by $1-\delta$.

Generally, however, the depth of the recursion can go beyond $O(\log |V|)$. 
This is because the choice of $\bar{v}$ is not uniformly at random, and there is a chance that one of $V_{-\bar{v}}$ or $V_{+\bar{v}}$ has almost the same size as $V$ with high probability.
To resolve this issue, we sample $\bar{v}$ from a set $U_d$ of vertices $v \in V$ such that  $|V_{-v}| \leq d$ and $|V_{+v}| \leq d$, where $d$ is $|V|$ times some constant in $(0,1)$.
We can prove that such a vertex set still has a vertex $v$ such that $r(v)$ is equal to the optimal value of the original instance.
With this modification, we can bound the depth of the recursion by $O(\log |V|)$ and obtain the approximation ratio of $1-\delta$.

By contrast, the average sensitivity analysis is far more involved.
The main part of our analysis is to prove that the distribution of the pivot $\bar{v}$ does not change much on average by deleting one of the potentially missing sets.
Specifically, we bound the following average total variation distance:
\begin{align}
    \frac{1}{n}\sum_{i=1}^{n}\TV(\Call{Alg}{V},\Call{Alg}{V\setminus S_i}),\label{eq:TValg}
\end{align}
where $\TV(\mathcal{X}_1,\mathcal{X}_2)$ represents the total variation distance of two output distributions $\mathcal{X}_1$ and $\mathcal{X}_2$.

Let us assume for now that the average total variation distance is small.
To bound the average sensitivity of our algorithm, for each $i \in \{1,2,\ldots,n\}$, we transport probability mass of $\Call{Alg}{V}$ corresponding to a particular choice of the pivot $\bar{v}$ to that of $\Call{Alg}{V\setminus S_i}$ corresponding to the same $\bar{v}$ as far as possible\footnote{When we bound the earth mover's distance from above, we can use any joint distribution $\mathcal{D}$ because we take the minimum over all possible joint distributions in~\eqref{eq:earth-mover}. In our analysis, we often construct $\mathcal{D}$ by specifying how we transport probability mass from one distribution to the other.}
In this way, we can transport $1-\TV(\Call{Alg}{V},\Call{Alg}{V\setminus S_i})$ amount of probability mass for each $i$.
For the probability mass transported this way, we recursively transport probability mass from $\Call{Alg}{V_{-\bar{v}}}$ to $\Call{Alg}{V_{-\bar{v}}\setminus S_i}$ and from $\Call{Alg}{V_{+\bar{v}}}$ to $\Call{Alg}{V_{+\bar{v}}\setminus S_i}$, and then apply the same analysis.
The remaining probability mass of $\TV(\Call{Alg}{V},\Call{Alg}{V\setminus S_i})$ is transported arbitrarily.
Its contribution to the average sensitivity can be bounded by $\TV(\Call{Alg}{V},\Call{Alg}{V\setminus S_i}) \cdot n$, which is small.

We now explain how we bound~\eqref{eq:TValg}.
An important observation is that when we delete a potentially missing set uniformly at random, the value $r(v)$ decreases by at most $\frac{r(v)}{n}$ in expectation for every $v \in V$.
Then one may think that if the factor $c$ is chosen appropriately, the decrease of the probability to select a particular $v$ as a pivot is small in expectation.
However, this idea does not work.
The main reason for this is that we sample a pivot from $U_d$, not $V$.
When $|V|$ decreases a lot by deleting a potentially missing set, a large number of vertices may join $U_d$, and the probability of choosing a vertex as a pivot may drastically change.

To resolve this issue, we sample the threshold $d$ from an interval, e.g., $[\frac{1}{2}|V|,\frac{3}{4}|V|]$.
Then, we analyze the average sensitivity by transporting the probability mass of $\Call{Alg}{V}$ corresponding to a particular choice of the threshold to that of $\Call{Alg}{V\setminus S_i}$ corresponding to the same threshold.
Note that we can do so, except for the probability mass that the threshold is in $\Call{Alg}{V}$ is in $[\frac{3}{4}|V\setminus S_i|,\frac{3}{4}|V|]$.
However, the average of this mass over $i$ is small and does not contribute much to the average sensitivity.

Similar issues arise when we choose the scaling factor $c$ and the value $\epsilon$ because they depend on $|V|$.
We can also resolve them by sampling these values from some intervals instead of fixing these values uniquely.


\paragraph{RNA Folding.}
Here, we describe the intuition behind our reduction from the RNA folding problem to the MWC problem.
Our reduction is inspired by a pseudo-polynomial time algorithm for solving constrained knapsack problems on trees~\cite{kumabe2019linear}, in which they reduced the dependency of the time complexity on the weight limit from quadratic to linear.
Before getting into details, we formally define the RNA folding problem.
\begin{prb}[RNA folding~\cite{RNA_nussinov}]
Let $A$ be a string of length $n$ over the alphabet $\Sigma$. Let $\mathcal{R}\subseteq \Sigma\times \Sigma$ be a binary relation. Two pairs of indices $(l,r)$ and $(l',r')$ are \emph{pseudoknot} if exactly one of $l'$ and $r'$ is located between $l$ and $r$, exclusively. The output is a set of pairs of indices $\{(l_1,r_1),\dots,(l_t,r_t)\}$ such that any two of $l_1,r_1,\dots, l_t,r_t$ are distinct. The goal is to maximize $t$ subject to $l_i < r_i$, $(A_{l_i},A_{r_i})\in \mathcal{R}$ for all $i\in \{1,\dots, t\}$, and no two different pairs in the output form a pseudoknot.
\end{prb}
In this work, the average sensitivity of an algorithm $\Call{Alg}{}$ for the RNA folding problem is defined as $\frac{1}{n}\sum_{i=1}^n d_{\mathrm{EM}}(\Call{Alg}{A}, \Call{Alg}{A^i})$, where for $i \in \{1,2,\ldots,n\}$, $A^i = (a_1,\dots, a_{i-1},a_{i+1},\dots a_n)$ is the sequence obtained from $A$ by dropping $a_i$ and the distance between two solutions used in the earth mover's distance is the size of their symmetric difference.


Let $A$ be an instance of the RNA folding.
We want to construct a transitive DAG $G$ such that a chain in $G$ corresponds to a solution for $A$ of the same weight and vice versa.

First, we observe that a feasible solution $X = \{(l_1,r_1),\dots, (l_t,r_t)\}$ for an instance $A$ of the RNA folding problem defines the transitive closure of a forest $T_X$ as follows.
We introduce a vertex in $T_X$ for each pair in $X$.
Then, we add an edge from $(l,r) \in X$ to another $(l',r') \in X$ whenever $[l',r']\subsetneq [l,r]$.
The resulting graph is the transitive closure of a forest because the feasible solution does not have a pseudoknot.
For the sake of analysis, we introduce a root vertex in $T_X$ corresponding to a pair $(0,n+1)$ and regard $T_X$ as a rooted tree.

For a feasible solution $X$ and $(l_i,r_i)\in X$, let $P_{X,i}$ be a path in $T_X$ from the root to the vertex $(l_i,r_i)$.
Let us consider a graph $G'$ whose vertex set is the set of all possible paths $\{P_{X,i}\}_{X,i}$.
We introduce only one vertex even if the same path arises from different feasible solutions.
For each feasible solution $X$ and a pair of base pairs $(l_i,r_i), (l_{i'},r_{i'})$ in $X$, we introduce an edge from $P_{X,i}$ to $P_{X,i'}$ if in some fixed pre-order transversal of $T_X$, $(l_i,r_i)$ appears earlier than $(l_{i'},r_{i'})$.
We can then show that the resulting graph $G'$ becomes acyclic and transitive if the pre-order transversals are consistent among $X$'s in a certain sense, and each chain in $G$ corresponds to a feasible solution for the original instance and vice versa.

An issue here is that the size of $G'$ is exponentially large, and thus Theorem~\ref{thm:main} applied on $G$ gives a polynomial bound on the average sensitivity, which is trivial.
To resolve this issue, we consider the \emph{heavy-light decomposition} of $T_X$.
An edge $(u,v)$ in $T_X$, where $v$ is a child of $u$, is \emph{heavy} if the size of the subtree rooted at $v$ is the maximum among those of all children of $u$. Otherwise, it is \emph{light}.
Ties are broken arbitrarily so that each non-leaf vertex in $T_X$ has exactly one heavy child.
Then, we construct a graph $G$ as follows.
For each feasible solution $X$ and $(l_i,r_i)\in X$, let $Q_{X,i}$ be the list of all light edges in the path $P_{X,i}$.
The vertex set of $G$ is the set of all possible lists $\{Q_{X,i}\}_{X,i}$.
We introduce only one vertex even if the same list arises from different feasible solutions.
For each feasible solution $X$ and a pair of base pairs $(l_i,r_i), (l_{i'},r_{i'})$ in $X$, we introduce an edge from $Q_{X,i}$ to $Q_{X,i'}$ if in some fixed pre-order transversal of $T_X$, $(l_i,r_i)$ appears earlier than $(l_{i'},r_{i'})$.
An important observation here is that because there are at most $\log n$ light edges on a path in $T_X$ and hence in $Q_{X,i}$, the size of $V(G)$ is bounded by $n^{O(\log n)}$. Therefore, by applying Theorem~\ref{thm:main} on $G$, we obtain a polylogarithmic average sensitivity bound.

\subsection{Organization}

The rest of this paper is organized as follows.
In Section~\ref{sec:overview}, we introduce our algorithm for the MWC problem and analyze its approximation ratio and average sensitivity.
In Section~\ref{sec:application}, we discuss some DP problems for which we can directly obtain stable-on-average algorithms by reducing to the MWC problem\@.
Finally, in Section~\ref{sec:RNA}, we show a stable-on-average algorithm for the RNA folding problem.


\section{Stable-on-average Algorithm for the Maximum Weight Chain Problem}\label{sec:overview}


In this section, we prove Theorem~\ref{thm:main}.
We describe our algorithm and show its basic properties in Section~\ref{subsec:algorithm-basic-properties}.
We analyze the approximation ratio and average sensitivity in Sections~\ref{sec:approx} and~\ref{sec:proofstrategy}, respectively.
The proof of a key technical lemma (Lemma~\ref{lem:pivottvd}) in the analysis of average sensitivity is provided in Section~\ref{sec:pivot}.

\subsection{Algorithm Description and Basic Properties}\label{subsec:algorithm-basic-properties}

Let $G=(V,E,w)$ be a transitive DAG with a vertex weight function $w:V \to \mathbb{R}_+$.
For a vertex set $U\subseteq V$ and a vertex $v\in U$, let $U_{-v} \subseteq V$ (resp., $U_{+v} \subseteq V$) be the set of vertices $u$ such that there is an edge from $u$ to $v$ (resp., from $v$ to $u$).
Note that owing to the transitivity of $G$, if there is a chain from $u$ to $v$, then there is an edge from $u$ to $v$, and hence $u \in U_{-v}$ holds.
Let $S_1,\dots, S_n$ be potentially missing antichains of $G$ such that each vertex in $V$ is contained in at least one and at most $K$ of them.
Because a chain cannot have two or more vertices from the same $S_i$, the size of any chain in $G$ is at most $n$.
Let $U \subseteq V$ be a vertex set.
Then, let $w(U)=\sum_{v\in U}w(v)$, $G[U]$ be the subgraph of $G$ induced by $U$, and $\OPT(U)$ be the maximum weight of a chain in $G[U]$.

\begin{algorithm}[t!]
\caption{Stable-on-average algorithm for the maximum weight chain problem}\label{alg:main}
\Procedure{\emph{\Call{Rec}{$U,\epsilon$}}}{
    Sample $c$ uniformly from  $\left[\frac{\epsilon \OPT(U)}{\log\left(|U|\epsilon^{-1}\right)},2\cdot \frac{\epsilon \OPT(U)}{\log\left(|U|\epsilon^{-1}\right)} \right]$\;
    Sample $d$ uniformly from $\left[\frac{1}{2}|U|,\frac{3}{4}|U|\right]$\;
    \If{$U=\emptyset$}{
        \Return $\emptyset$\;
    }
    \For{$v\in U$}{
        Let $r(v)$ be the maximum weight of a chain that includes $v$\;\label{line:definerv}
    }
    Let $U_d$ be the set of vertices $v \in U$ with $\max\left(|U_{-v}|,|U_{+v}|\right)\leq d$\label{line:Ud}\;
    Sample $\bar{v}\in U_d$ with probability proportional to $\exp(r(\bar{v})/c)$\label{line:main-exponential-mechanism}\;
    \Return \Call{Rec}{$U_{-\bar{v}},\epsilon$} $\cup$ $\{\bar{v}\}$ $\cup$ \Call{Rec}{$U_{+\bar{v}},\epsilon$}\;
}
\Procedure{\emph{\Call{MWC}{$G=(V,E,w), \delta$}}}{
    Sample $\epsilon^{-1}$ uniformly from $\left[17\delta^{-1}\log(|V|),34\delta^{-1}\log(|V|)\right]$\;
    \Return \Call{Rec}{$V,\epsilon$}\; 
}
\end{algorithm}



Our algorithm is given in Algorithm~\ref{alg:main}.
Given a vertex set $U$, we select a vertex $v \in U$, which we call a \emph{pivot}, in a nearly optimal chain with respect to $G[U]$.
Then we recursively apply the algorithm on $U_{-v}$ and $U_{+v}$.
To bound the depth of the recursion, we select a vertex $v$ from $U_d$ (defined at Line~\ref{line:Ud}) so that both $|U_{-v}|$ and $|U_{+v}|$ are of at most a constant, say $\frac{3}{4}$, fraction of $|U|$.
Clearly, the running time is polynomial.

The following lemma ensures that an optimal chain has a vertex in $U_d$ for any $d\geq \frac{1}{2}|U|$.
\begin{lemma}
    For any $U \subseteq V$ and $d\geq \frac{1}{2}|U|$, there is a vertex $v \in U_d$ with $r(v)=\OPT(U)$, where $r(v)$ is as defined at Line~\ref{line:definerv} of Algorithm~\ref{alg:main}. In particular, $\max_{v\in U_d}r(v)=\OPT(U)$.
\end{lemma}
\begin{proof}
Let $P=(v_1,\dots, v_k)$ be a maximal chain that attains $\OPT(U)$.
Let $i$ be the last index with $|U_{-v_i}|\leq \frac{|U|}{2}$. We prove $|U_{+v_i}|\leq \frac{|U|}{2}$.

If $i=k$, then $|U_{+v_i}|=0$ because $P$ is maximal, and we are done.
Otherwise, we have
\begin{align*}
    |U_{+v_i}|&=|U_{+v_i}\setminus U_{-v_{i+1}}|+|U_{+v_i}\cap U_{-v_{i+1}}|
    =|U_{+v_i}\setminus U_{-v_{i+1}}|
    \leq |U\setminus U_{-v_{i+1}}|\leq \frac{|U|}{2},
\end{align*}
where the second equality is from $|U_{+v_i}\cap U_{-v_{i+1}}|=0$ that is obtained from the maximality of $P$, and the last inequality is from the definition of $i$.
\end{proof}

For a vertex set $U$ and an index $i$, let $U^i=U\setminus S_i$. The following lemma is useful in our analysis.
\begin{lemma}\label{lem:sumdiff}
We have
\begin{align}
    \sum_{i=1}^{n}(\OPT(U)-\OPT(U^i))\leq K\OPT(U).
\end{align}
\end{lemma}
\begin{proof}
Let $P$ be a chain that attains $\OPT(U)$. Then, we have
\begin{align*}
    \sum_{i=1}^{n}(\OPT(U)-\OPT(U^i))
    \leq \sum_{i=1}^{n}\left(w(P)-w(P\setminus S_i)\right)
    = \sum_{i=1}^{n}w(P\cap S_i)\leq K\cdot w(P) = K\OPT(U),
\end{align*}
where the first inequality is from the fact that $P\setminus S_i$ is a feasible solution and the second inequality is from the fact that each vertex in $P$ belong to at most $K$ of $S_1,\dots, S_n$.
\end{proof}

Throughout the paper, for a distribution $X$ and a condition $P$, we denote the conditional distribution of $\mathcal{X}$ conditioned on $P$ by $(\mathcal{X} \mid P)$.
The following lemma is useful in our analysis.
The proof is given in Appendix~\ref{app:app}. 
\begin{lemma}\label{lem:benri}
Let $\D(\cdot,\cdot)$ denote either the earth mover's distance or the total variation distance.
Let $\Call{Alg}{}$ be a randomized algorithm. Suppose there is a parameter $p$ (resp., $p^i$) used in $\Call{Alg}{}$ for the instance $U$ (resp., $U^i$), sampled from the uniform distribution over $[B,(1+t)B]$ (resp., $[B^i,(1+t)B^i]$).
Let $M$ be an upper bound of $\D(\Call{Alg}{U},\Call{Alg}{U^i})$.
Then for any $t>0$, we have
\begin{align*}
    &\frac{1}{n}\sum_{i=1}^{n}\D(\Call{Alg}{U},\Call{Alg}{U^i})\\
    &\leq \frac{1}{tB}\int_{B}^{(1+t)B}\left(\frac{1}{n}\sum_{i=1}^{n}\D\left((\Call{Alg}{U}\mid p=\hat{p}),(\Call{Alg}{U^i}\mid p^i=\hat{p})\right)\right)\mathrm{d}\hat{p}+\frac{M}{n}\cdot \frac{1+t}{t}\cdot  \sum_{i=1}^{n}\left|1-\frac{B^i}{B}\right|.
\end{align*}
\end{lemma}

\subsection{Approximation Ratio}\label{sec:approx}

In this section, we analyze the approximation ratio of Algorithm~\ref{alg:main}.
First we analyze the loss caused by the exponential mechanism at Line~\ref{line:main-exponential-mechanism}.

\begin{lemma}\label{lem:expapprox}
Let $U \subseteq V$ and $\bar{v}$ be as defined in $\Call{Rec}{}$. Then, we have
\[
    \E[r(\bar{v})]\geq (1-2\epsilon) \OPT(U).
\]
\end{lemma}
\begin{proof}
We have
\begin{align*}
    & \Pr[r(\bar{v}) \leq (1-\epsilon)\OPT(U)]
    = \frac{\sum_{v\in U_d: r(v)\leq (1-\epsilon) \OPT(U)}\exp(r(v)/c)}{\sum_{v\in U_d}\exp(r(v)/c)}
    \leq \frac{|U_d|\exp((1-\epsilon)\OPT(U)/c)}{\sum_{v\in U_d}\exp(r(v)/c)} \\
    & \leq \frac{|U_d|\exp((1-\epsilon)\OPT(U)/c)}{\exp(\OPT(U)/c)}
    = |U_d|\exp(-\epsilon \cdot \OPT(U)/c)
    = |U_d|\cdot \frac{\epsilon}{|U|}
    \leq \epsilon,
\end{align*}
where the first inequality is from the algorithm and the last equality is from the definition of $c$.
Therefore, we have
\[
    \E[r(\bar{v})]
    \geq (1-\epsilon)(1-\epsilon)\OPT(U)
    \geq (1-2\epsilon)\OPT(U).
    \qedhere
\]
\end{proof}

Next, we analyze the loss caused by recursion and complete the analysis of approximation ratio.
\begin{lemma}\label{lem:approx}
For any $U \subseteq V$ and $\epsilon > 0$, we have
\[
    \E\left[w(\Call{Rec}{U,\epsilon})\right]\geq (1-17\epsilon\log |U|)\OPT(U).
\]
\end{lemma}
\begin{proof}
We prove by induction on $|U|$.
The statement clearly holds when $|U|=1$.

Suppose $|U|>1$.
Let $\bar{v}$ be as defined in $\Call{Rec}{U}$.
Then, we have
\begin{align*}
    \E\left[w(f(U))\right]
    &= \E\left[\sum_{v\in U_d}\Pr[v=\bar{v}]\left(\E\left[w(\Call{Rec}{U_{-v}})\right]+w(v)+\E\left[w(\Call{Rec}{U_{+v}})\right]\right)\right]\\
    &\geq \E\left[\sum_{v\in U_d}\Pr[v=\bar{v}]\left((1-17\epsilon \log |U_{-v}|)\OPT(U_{-v})+w(v)+(1-17\epsilon \log |U_{+v}|)\OPT(U_{+v})\right)\right]\\
    &\geq \left(1-17\epsilon \log \left(\frac{3}{4}|U|\right)\right) \E\left[\sum_{v\in U_d}\Pr[v=\bar{v}]\left(\OPT(U_{-v})+w(v)+\OPT(U_{+v})\right)\right]\\
    &\geq \left(1-17\epsilon \log \left(\frac{3}{4}|U|\right)\right) \E\left[\sum_{v\in U_d}\Pr[v=\bar{v}]r(v)\right]\\
    &= \left(1-17\epsilon \log \left(\frac{3}{4}|U|\right)\right) \E\left[r(\bar{v})\right]\\
    &\geq \left(1-17\epsilon \log \left(\frac{3}{4}|U|\right)\right)(1-2\epsilon)\OPT(U)\\
    &\geq \left(1-17\epsilon\log \left(\frac{3}{4}|U|\right)-2\epsilon \right)\OPT(U)\\
    &\geq (1-17\epsilon\log |U|)\OPT(U),
\end{align*}
where the first inequality is from the induction hypothesis, the third inequality is from the definition of $r(v)$, the fourth inequality is from Lemma~\ref{lem:expapprox}, and the last inequality is from $17\log\left(\frac{4}{3}\right)\geq 2$.
\end{proof}

\subsection{Average Sensitivity}\label{sec:proofstrategy}

In this section, we give an average sensitivity bound of Algorithm~\ref{alg:main} and complete the proof of Theorem~\ref{thm:main}.


To evaluate the earth mover's distance in~\eqref{eq:avg-sensitivity}, we consider transporting probability mass of \Call{MWC}{$V$} corresponding to a particular choice of $\epsilon$ to that of \Call{MWC}{$V\setminus S_i$} corresponding to the same choice of $\epsilon$.

First, we focus on analyzing the average sensitivity of the procedure \Call{Rec}{} for a fixed $\epsilon$, which is defined by
\begin{align*}
    \frac{1}{n}\sum_{i=1}^{n}\EM(\Call{Rec}{V,\epsilon}, \Call{Rec}{V\setminus S_i,\epsilon}).
\end{align*}
As $\epsilon$ is fixed, we drop it from the argument below.
We transport probability mass of \Call{Rec}{$V$} corresponding to a particular choice of the pivot $\bar{v}$ to that of \Call{Rec}{$V\setminus S_i$} corresponding to the same pivot as far as possible.
For a fixed $\bar{v}$, we transport probability mass from \Call{Rec}{$V_{-\bar{v}}$} (resp., \Call{Rec}{$V_{+\bar{v}}$}) to \Call{Rec}{$V_{-\bar{v}}\setminus S_i$} (resp., \Call{Rec}{$V_{+\bar{v}}\setminus S_i$}) as far as possible, where the mass is transported recursively in the same way as was done from \Call{Rec}{$V$} to \Call{Rec}{$V\setminus S_i$}.
Because \Call{Rec}{$V_{-\bar{v}}$} and \Call{Rec}{$V_{+\bar{v}}$} (resp., \Call{Rec}{$V_{-\bar{v}}\setminus S_i$} and \Call{Rec}{$V_{+\bar{v}}\setminus S_i$}) are independent for fixed $\bar{v}$, we obtain a probability transportation from the probability mass of \Call{Rec}{$V$} to that of \Call{Rec}{$V\setminus S_i$} as their direct product.
The remaining probability mass is transported arbitrarily.

For $U\subseteq V$, we denote the random variable $\bar{v}$ chosen in $\Call{Rec}{U}$ by $\bar{v}(U)$. Let $n_U$ be the number of potentially missing sets $S_i$ with $U\cap S_i\neq \emptyset$.
The main part of our analysis is to bound the average total variation distance of the pivot. 
Specifically, we prove the following.
\begin{lemma}\label{lem:pivottvd}
For any fixed $0<\epsilon\leq 0.05$, we have
\begin{align}
    \frac{1}{n_U}\sum_{i\colon U\cap S_i\neq \emptyset}\TV\left(\bar{v}(U),\bar{v}(U\setminus S_i)\right)\leq O\left(\frac{1}{n_U}\cdot K\epsilon^{-1}\log\left(|U|\epsilon^{-1}\right)\right).\label{eq:pivottvd}
\end{align}
\end{lemma}
We postpone the proof of Lemma~\ref{lem:pivottvd} to Section~\ref{sec:pivot} and continue our discussion assuming that Lemma~\ref{lem:pivottvd} holds.

For each $i\in \{1,\dots, n\}$ with $U\cap S_i\neq \emptyset$, we transport the probability mass from $\Call{Rec}{U}$ to $\Call{Rec}{U\setminus S_i}$ as far as possible.
Using our transportation scheme, the total amount of probability mass that is transported to that with a different pivot is $\TV\left(\bar{v}(U),\bar{v}(U\setminus S_i)\right)$.
For this mass, we bound the Hamming distance between solutions on $U$ and $U\setminus S_i$ by $n_U$, which is a trivial upper bound from the fact that $U$ is covered by $n_U$ antichains.
By taking the average over $i$, we have that this bound contributes to the average sensitivity by at most $\eqref{eq:pivottvd}\cdot n_U=O\left(K\epsilon^{-1}\log\left(|U|\epsilon^{-1}\right)\right)$.

Next, we focus on the mass transported to that with the same pivot.
Here, we must also analyze the average sensitivity incurred by recursions.
For an integer $j \geq 0$, let $\mathcal{U}_j$ be the family of sets $U \subseteq V$ such that $\Call{Rec}{U}$ is called in one of the recursion steps of depth $j$.
Here, we regard $\Call{Rec}{V}$ as the unique recursion step of depth $0$, and hence $\mathcal{U}_0=\{V\}$.
Because we choose the pivot from $U_d$, we have $|U|\leq \left(\frac{3}{4}\right)^j|V|$ for all $U\in \mathcal{U}_j$, and it follows that the maximum integer $j$ with $\mathcal{U}_j\neq \emptyset$, denoted $k$, is $O(\log |V|)$.
Now the average sensitivity of $\Call{Rec}{}$ is bounded by
\begin{align}
    \frac{1}{n}\sum_{i=1}^{n} \E\left[\sum_{j=0}^{k} \sum_{U\in \mathcal{U}_j, S_i\cap U\neq \emptyset}\TV\left(\bar{v}(U),\bar{v}(U\setminus S_i)\right)\cdot n_U\right].\label{eq:recemd}
\end{align}
For any $i\in \{1,\dots, n\}$, at least one of $U_{-\bar{v}}\cap S_i$ and $U_{+\bar{v}}\cap S_i$ is empty because $S_i$ is an antichain.
Therefore for every $j \in \{0,\ldots,k\}$, there is at most one set $U\in \mathcal{U}_j$ with $S_i\cap U\neq \emptyset$, which implies that the third summation in~\eqref{eq:recemd} is taken over at most one set.

We have the following main lemma.
\begin{lemma}\label{lem:main}
We have
\[
    \frac{1}{n}\sum_{i=1}^{n}\EM\left(\Call{Rec}{V}, \Call{Rec}{V\setminus S_i}\right)\leq O\left(K\epsilon^{-1}\log |V|\log\left(|V|\epsilon^{-1}\right)\right).
\]
\end{lemma}
\begin{proof}
We have
\begin{align*}
    \frac{1}{n}\sum_{i=1}^{n}\EM\left(\Call{Rec}{V}, \Call{Rec}{V\setminus S_i}\right)
    &\leq \frac{1}{n}\sum_{i=1}^{n}\E\left[\sum_{j=0}^{k}\sum_{U\in \mathcal{U}_j, S_i\cap U\neq \emptyset}\TV\left(\bar{v}(U),\bar{v}(U\setminus S_i)\right)\cdot n_U\right] \\
    &= \sum_{j=0}^{k}\E\left[\frac{1}{n}\sum_{U\in \mathcal{U}_j}\left(\sum_{i\colon S_i\cap U\neq\emptyset}\TV\left(\bar{v}(U),\bar{v}(U\setminus S_i)\right)\cdot n_U\right)\right]\\
    &\leq \sum_{j=0}^{k}\E\left[\frac{1}{n}\sum_{U\in \mathcal{U}_j}O\left(K\epsilon^{-1}\log\left(|U|\epsilon^{-1}\right)\right)\cdot n_U\right]\\
    &\leq \sum_{j=0}^{k}O\left(K\epsilon^{-1}\log\left(|V|\epsilon^{-1}\right)\right)\\
    &\leq O\left(K\epsilon^{-1}\log |V|\log\left(|V|\epsilon^{-1}\right)\right),
\end{align*}
where the first inequality is from~\eqref{eq:recemd}, the second inequality is from Lemma~\ref{lem:pivottvd}, the third inequality is from
$\sum_{U\in \mathcal{U}_j}n_U\leq n$ and $|U|\leq |V|$, and the last inequality is from $k\leq O(\log |V|)$.
\end{proof}


We analyze the average sensitivity of \Call{MWC}{}:
\begin{proof}[Proof of Theorem~\ref{thm:main}]
If $\delta^{-1}>|V|$, then the theorem clearly holds because the average sensitivity is at most $|V| = O(K\delta^{-1}\log^3 |V|)$.
Therefore, we assume $\delta^{-1}\leq |V|$.
Then, from Lemma~\ref{lem:approx}, the approximation ratio is at most $1-\delta$.

Let $B=17\delta^{-1}\log |V|$. The average sensitivity is bounded as
\begin{align*}
    &\frac{1}{n}\sum_{i=1}^{n}\EM\left(\Call{MWC}{G,\delta},\Call{MWC}{G[V \setminus S_i],\delta}\right)\\
    &\leq \frac{1}{B}\int_{B}^{2B}\left(\frac{1}{n}\sum_{i=1}^{n}\EM\left(\left(\Call{MWC}{G,\delta})\mid \epsilon^{-1}=b\right),\left(\Call{MWC}{G[V \setminus S_i],\delta}\mid \epsilon^{-1}=b\right)\right)\right)\mathrm{d}b\\
    &\quad \quad +2\cdot \sum_{i=1}^{n}\left|1-\frac{\log(|V|-|S_i|)}{\log |V|}\right|\\
    &\leq \frac{1}{B}\int_{B}^{2B}O\left(Kb\log^2\left(|V|b\right)\right)\mathrm{d}b
    +2\cdot \sum_{i=1}^{n}\left(1-\frac{\log(|V|-|S_i|)}{\log |V|}\right)\\
    &\leq O\left(KB\log^2\left(|V|B\right)\right)+2\cdot \sum_{i=1}^{n}\left(1-\frac{\log(|V|-|S_i|)}{\log |V|}\right)\\
    &\leq O\left(KB\log^2\left(|V|B\right)\right)+2\cdot 1\cdot \frac{K|V|}{|V|-1}\\
    &\leq O\left(KB\log^2\left(|V|B\right)\right)+O(K)\\
    &\leq O\left(KB\log^2\left(|V|B\right)\right) = O(K\delta^{-1}\log |V|\log^2 \left(|V|\delta^{-1})\right)\leq O(K\delta^{-1}\log^3 |V|),
\end{align*}
where the first inequality is from Lemma~\ref{lem:benri}, the second inequality is from Lemma~\ref{lem:pivottvd}, the fourth inequality is from the convexity of $\log x$, and the last inequality is from $\delta^{-1}\leq |V|$.
\end{proof}

\subsection{Proof of Lemma~\ref{lem:pivottvd}}\label{sec:pivot}

In this section, we prove Lemma~\ref{lem:pivottvd}.
We focus on the case $U=V$ because the statement for $U\subseteq V$ is obtained by replacing $n$ by $n_U$ and the potentially missing sets $S_1,\dots, S_n$ by $S_{i_1}\cap U,\dots, S_{i_{n_U}}\cap U$, where $S_{i_1},\dots, S_{i_{n_U}}$ are the potentially missing sets with nonempty intersection with $U$.

Since Lemma~\ref{lem:pivottvd} is trivial when $|V| = 1$, we assume that $|V|\geq 2$.
In this section, we denote $V\setminus S_i$ by $V^i$ for notational simplicity.
We also denote $r(v)$ and $\bar{v}$ in $\Call{Rec}{V^i}$ by $r^i(v)$ and $\bar{v}^i$, respectively.
Furthermore, we assume that each potentially missing set is a proper subset of $V$. This assumption does not lose the generality because adding $V$ itself as a potentially missing set increases the LHS of~\eqref{eq:pivottvd} by at most $\frac{1}{n}$ and its RHS by at least $O\left(\frac{1}{n}\epsilon^{-1}\log\left(|V|\epsilon^{-1}\right)\right)$.

Now, we analyze the contribution of sampling parameters $c$ and $d$ in Algorithm~\ref{alg:main} to the total variation distance, using Lemma~\ref{lem:benri}.
We start by analyzing $c$.
\begin{lemma}
We have
\begin{align*}
    \frac{1}{n}\sum_{i=1}^{n}\TV(\bar{v},\bar{v}^i)
    \leq\sup_{\hat{c}\in [B,2B]}\left(\frac{1}{n}\sum_{i=1}^{n}\TV\left((\bar{v} \mid c = \hat{c}),(\bar{v}^i \mid c^i = \hat{c})\right)\right)+\frac{1}{n}\cdot 8K\log\left(|V|\epsilon^{-1}\right),
\end{align*}
where $B=\frac{\epsilon\OPT(V)}{\log\left(|V|\epsilon^{-1}\right)}$.
\end{lemma}
\begin{proof}
For $i=1,\dots, n$, let $B^i=\frac{\epsilon\OPT(V^i)}{\log(|V^i|\epsilon^{-1})}$. Then, by applying Lemma~\ref{lem:benri}, we obtain
\begin{align}
    &\frac{1}{n}\sum_{i=1}^{n}\TV(\bar{v},\bar{v}^i) \nonumber \\
    &\leq \frac{1}{2B}\int_{B}^{2B}\left(\frac{1}{n}\sum_{i=1}^{n}\TV\left((\bar{v} \mid c = \hat{c}),(\bar{v}^i \mid c^i = \hat{c})\right)\right)\mathrm{d}\hat{c}+\frac{1}{n}\cdot 2\cdot \sum_{i=1}^{n}\left|1-\frac{B^i}{B}\right| \nonumber \\
    &\leq\sup_{\hat{c}\in [B,2B]}\left(\frac{1}{n}\sum_{i=1}^{n}\TV\left((\bar{v} \mid c = \hat{c}),(\bar{v}^i \mid c^i = \hat{c})\right)\right)+\frac{1}{n}\cdot 2\cdot \sum_{i=1}^{n}\frac{1}{B}\left|B-B^i\right|.\label{eq:cerror1}
\end{align}
Now, we have
\begin{align}
    \frac{1}{B}\left|B-B^i\right|
    &=\frac{1}{B}\left|\frac{\epsilon\OPT(V^i)}{\log(|V^i|\epsilon^{-1})}-\frac{\epsilon\OPT(V)}{\log\left(|V|\epsilon^{-1}\right)}\right| \nonumber \\
    &\leq \frac{1}{B}\left|\frac{\epsilon\OPT(V^i)}{\log(|V^i|\epsilon^{-1})}-\frac{\epsilon\OPT(V)}{\log(|V^i|\epsilon^{-1})}\right|
    +\frac{1}{B}\left|\frac{\epsilon\OPT(V)}{\log(|V^i|\epsilon^{-1})}-\frac{\epsilon\OPT(V)}{\log\left(|V|\epsilon^{-1}\right)}\right|,\label{eq:cerror2}
\end{align}
where the inequality is from the triangle inequality.
Now, we have
\begin{align}
    \sum_{i=1}^{n}\frac{1}{B}\left|\frac{\epsilon\OPT(V^i)}{\log\left(|V^i|\epsilon^{-1}\right)}-\frac{\epsilon\OPT(V)}{\log(|V^i|\epsilon^{-1})}\right|
    &= \sum_{i=1}^{n}\frac{\log\left(|V|\epsilon^{-1}\right)}{\log\left(|V^i|\epsilon^{-1}\right)}\left(1-\frac{\OPT(V^i)}{\OPT(V)}\right) \nonumber \\
    &\leq \sum_{i=1}^{n}\frac{\log\left(|V|\epsilon^{-1}\right)}{\log\left(|V^i|\epsilon^{-1}\right)}\cdot 1 \nonumber \\
    &\leq \frac{K|V|}{|V|-1}\cdot \frac{\log\left(|V|\epsilon^{-1}\right)}{\log(\epsilon^{-1})} \nonumber \\
    &\leq 2K\log\left(|V|\epsilon^{-1}\right),\label{eq:cerror3}
\end{align}
where the last inequality is from $|V|\geq 2$ and $\epsilon < 0.2$. The second inequality is obtained as follows. Since $\sum_{i=1}^{n}(|V|-|V^i|) = \sum_{i=1}^{n}|S_i|\leq K|V|$, we have
\begin{align*}
    \sum_{i=1}^{n}\frac{1}{\log\left(|V^i|\epsilon^{-1}\right)}
    &\leq \max_{1\leq x_1,\ldots, x_n\leq |V|-1, x_1+\cdots+x_n\leq K|V|}\sum_{i=1}^{n}\frac{1}{\log\left((|V|-x_i)\epsilon^{-1}\right)}\\
    &\leq \frac{K|V|}{|V|-1}\cdot \frac{1}{\log(1\cdot \epsilon^{-1})},
\end{align*}
where the second inequality is from the convexity of $\frac{1}{\log((|V|-x)\epsilon^{-1})}$, when considered as a function of $x$.
Furthermore, we have
\begin{align}
    \sum_{i=1}^{n}\frac{1}{B}\left|\frac{\epsilon\OPT(V)}{\log(|V^i|\epsilon^{-1})}-\frac{\epsilon\OPT(V)}{\log\left(|V|\epsilon^{-1}\right)}\right|
    &= \sum_{i=1}^{n}\log\left(|V|\epsilon^{-1}\right)\left(\frac{1}{\log(|V^i|\epsilon^{-1})}-\frac{1}{\log\left(|V|\epsilon^{-1}\right)}\right) \nonumber \\
    &\leq \log\left(|V|\epsilon^{-1}\right)\cdot\left(\frac{K|V|}{|V|-1}\cdot \left(\frac{1}{\log(\epsilon^{-1})}-\frac{1}{\log\left(|V|\epsilon^{-1}\right)}\right)\right) \nonumber \\
    &\leq 2K\log\left(|V|\epsilon^{-1}\right),\label{eq:cerror4}
\end{align}
where the first inequality is obtained by the similar argument as above, using the convexity of $\frac{1}{\log((|V|-x)\epsilon^{-1})}-\frac{1}{\log(|V|\epsilon^{-1})}$ as a function of $x$, and the second inequality is from $|V|\geq 2$ and $\epsilon < 0.05$.

We obtain the claim by combining~\eqref{eq:cerror1},~\eqref{eq:cerror2},~\eqref{eq:cerror3}, and~\eqref{eq:cerror4}.
\end{proof}
Next, we evaluate the contribution of sampling $d$ to the total variation distance.
\begin{lemma}\label{lem:impact_cd}
We have
\begin{align*}
    &\frac{1}{n}\sum_{i=1}^{n}\TV(\bar{v},\bar{v}^i)\\
    &\leq\sup_{\hat{c}\in [B,2B]}\left(\frac{4}{|V|}\int_{\frac{1}{2}|V|}^{\frac{3}{4}|V|}\left(\frac{1}{n}\sum_{i=1}^{n}\TV\left((\bar{v} \mid c = \hat{c},d=\hat{d}),(\bar{v}^i \mid c^i = \hat{c},d^i=\hat{d})\right)\right)\mathrm{d}\hat{d}\right)+\frac{1}{n}\cdot 9K\log\left(|V|\epsilon^{-1}\right),
\end{align*}
where $B=\frac{\epsilon\OPT(V)}{\log\left(|V|\epsilon^{-1}\right)}$.
\end{lemma}
\begin{proof}
We have
\begin{align*}
    &\frac{1}{n}\sum_{i=1}^{n}\TV((\bar{v} \mid c = \hat{c}),(\bar{v}^i \mid c^i = \hat{c}))\\
    &\leq \frac{4}{|V|}\int_{\frac{1}{2}|V|}^{\frac{3}{4}|V|}\left(\frac{1}{n}\sum_{i=1}^{n}\TV\left((\bar{v} \mid c = \hat{c},d=\hat{d}),(\bar{v}^i \mid c^i = \hat{c},d^i=\hat{d})\right)\right)\mathrm{d}\hat{d}+\frac{1}{n}\cdot 3\cdot \sum_{i=1}^{n}\left|1-\frac{\frac{1}{2}|V^i|}{\frac{1}{2}|V|}\right|\\
    &\leq \frac{4}{|V|}\int_{\frac{1}{2}|V|}^{\frac{3}{4}|V|}\left(\frac{1}{n}\sum_{i=1}^{n}\TV\left((\bar{v} \mid c = \hat{c},d=\hat{d}),(\bar{v}^i \mid c^i = \hat{c},d^i=\hat{d})\right)\right)\mathrm{d}\hat{d}+\frac{1}{n}\cdot 3K,
\end{align*}
where the first inequality is from Lemma~\ref{lem:benri}, and the second inequality is from the fact that each vertex in $V$ is contained in at most $K$ of the potentially missing sets.
Therefore, we have
\begin{align*}
    &\frac{1}{n}\sum_{i=1}^{n}\TV(\bar{v},\bar{v}^i)\\
    &\leq\sup_{\hat{c}\in [B,2B]}\left(\frac{1}{n}\sum_{i=1}^{n}\TV\left((\bar{v} \mid c = \hat{c}),(\bar{v}^i \mid c^i = \hat{c})\right)\right)+\frac{1}{n}\cdot 8K\log\left(|V|\epsilon^{-1}\right)\\
    &\leq\sup_{\hat{c}\in [B,2B]}\left(\frac{4}{|V|}\int_{\frac{1}{2}|V|}^{\frac{3}{4}|V|}\left(\frac{1}{n}\sum_{i=1}^{n}\TV\left((\bar{v} \mid c = \hat{c},d=\hat{d}),(\bar{v}^i \mid c^i = \hat{c},d^i=\hat{d})\right)\right)\mathrm{d}\hat{d}+\frac{1}{n}\cdot 3K\right) \\
    & \qquad+\frac{1}{n}\cdot 8K\log\left(|V|\epsilon^{-1}\right)\\
    &\leq\sup_{\hat{c}\in [B,2B]}\left(\frac{4}{|V|}\int_{\frac{1}{2}|V|}^{\frac{3}{4}|V|}\left(\frac{1}{n}\sum_{i=1}^{n}\TV\left((\bar{v} \mid c = \hat{c},d=\hat{d}),(\bar{v}^i \mid c^i = \hat{c},d^i=\hat{d})\right)\right)\mathrm{d}\hat{d}\right) \\
    & \qquad +\frac{1}{n}\cdot 9K\log\left(|V|\epsilon^{-1}\right).
    \qedhere
\end{align*}
\end{proof}

Now, we focus on bounding the value
\begin{align*}
    \frac{1}{n}\sum_{i=1}^{n}\TV\left((\bar{v} \mid c = \hat{c},d=\hat{d}),(\bar{v}^i \mid c^i = \hat{c},d^i=\hat{d})\right)
\end{align*}
for fixed $\hat{c}$ and $\hat{d}$.
For notational simplicity, we write conditional probabilities such as $\Pr[\bar{v}=v \mid c=\hat{c}, d=\hat{d}]$ and $\Pr[\bar{v}=v^i \mid c^i=\hat{c}, d^i=\hat{d}]$ as $\Pr[\bar{v}=v \mid \hat{c},\hat{d}]$ and $\Pr[\bar{v}=v^i \mid \hat{c},\hat{d}]$, respectively.
Now, we have
\begin{align}
    &\frac{1}{n}\sum_{i=1}^{n}\TV\left((\bar{v} \mid c = \hat{c},d=\hat{d}),(\bar{v}^i \mid c^i = \hat{c},d^i=\hat{d})\right) \nonumber \\
    &= \frac{1}{n}\sum_{i=1}^{n}\sum_{v\in S_i}\Pr[\bar{v}=v|\hat{c},\hat{d}] + \frac{1}{n}\sum_{i=1}^{n}\sum_{v\in V_d\setminus S_i}\max\left(0,\Pr[\bar{v}=v|\hat{c},\hat{d}]-\Pr[\bar{v}^i=v|\hat{c},\hat{d}]\right)\label{eq:sumprob}.
\end{align}
In the second term, because $\max\left(0,\Pr[\bar{v}=v\mid\hat{c},\hat{d}]-\Pr[\bar{v}^{u}=v\mid\hat{c},\hat{d}]\right)$ is positive only if $v\in V_d$, we can take the sum over $v\in V_d\setminus S_i$ instead of $v\in V\setminus S_i$.

We evaluate the two terms of~\eqref{eq:sumprob} separately. The first term is simple.
\begin{lemma}\label{lem:deletev}
For any $\hat{c}$ and $\hat{d}$, we have
\[
    \frac{1}{n}\sum_{i=1}^{n}\sum_{v\in S_i}\Pr[\bar{v}=v\mid\hat{c},\hat{d}]\leq \frac{1}{n}\cdot K.
\]
\end{lemma}
\begin{proof}
We have
\[
    \frac{1}{n}\sum_{i=1}^{n}\sum_{v\in S_i}\Pr[\bar{v}=v\mid \hat{c},\hat{d}]=\frac{1}{n}\sum_{i=1}^{n}\Pr[\bar{v}\in S_i\mid\hat{c},\hat{d}]\leq \frac{1}{n}\cdot K,
\]
where the inequality is from the assumption that each element of $V$ is contained in at most $K$ of $S_1,\dots, S_n$.
\end{proof}

Let us evaluate the second term. Now, for $v\in V_d\setminus S_i$, we have
\begin{align}
    &\max\left(0,\Pr[\bar{v}=v\mid\hat{c},\hat{d}]-\Pr[\bar{v}^i=v\mid\hat{c},\hat{d}]\right) \nonumber
    = \max\left(0,\frac{\exp(r(v)/c)}{\sum_{v'\in V_{d}}\exp(r(v')/c)}-\frac{\exp(r^i(v)/c)}{\sum_{v'\in V_{d}^i}\exp(r^i(v')/c)}\right) \nonumber \\
    &\leq \frac{\exp(r(v)/c)-\exp(r^i(v)/c)}{\sum_{v'\in V_{d}}\exp(r(v')/c)}+\max\left(0,\frac{\exp(r^i(v)/c)}{\sum_{v'\in V_{d}}\exp(r(v')/c)}-\frac{\exp(r^i(v)/c)}{\sum_{v'\in V_{d}^i}\exp(r^i(v')/c)}\right)\label{eq:pseudotriangle},
\end{align}
where the equality is from the design of the algorithm and the inequality is from the following inequality
\[
    \max(0,b-a)\leq (b-x)+\max(0,x-a),
\]
which holds for any $x\leq b$.

Let us give some intuition about the two terms in~\eqref{eq:pseudotriangle}.
Consider deleting $S_i$. The first term of~\eqref{eq:pseudotriangle} represents the decrease of the value $\Pr[\bar{v}=v\mid\hat{c},\hat{d}]$ caused by the decrease of $r(v)$, which is positive when $S_i$ crosses all MWCs $v$. The second term of~\eqref{eq:pseudotriangle} represents the decrease of the value $\Pr[\bar{v}=v\mid\hat{c},\hat{d}]$ caused by the increase of the denominator $\sum_{v\in V_{d}}\exp(r(v)/c)$, which happens when $V_{d}\subsetneq V_{d}^i$ holds.

The next lemma bounds the first term of~\eqref{eq:pseudotriangle}.
\begin{lemma}\label{lem:triangle_1}
Conditioned on having chosen $c=c^i=\hat{c}$ and $d=d^i=\hat{d}$, we have
\[
    \frac{1}{n}\sum_{i=1}^{n}\sum_{v\in V_{d}\setminus S_i}\frac{\exp(r(v)/c)-\exp(r^i(v)/c)}{\sum_{v'\in V_{d}}\exp(r(v')/c)}\leq \frac{1}{n}\cdot K\epsilon^{-1}\log\left(|V|\epsilon^{-1}\right).
\]
\end{lemma}
\begin{proof}
We have
\begin{align}
    \frac{\exp(r(v)/c)-\exp(r^i(v)/c)}{\sum_{v'\in V_{d}}\exp(r(v')/c)}
    &= \Pr[\bar{v}=v\mid\hat{c},\hat{d}]\left(1-\frac{\exp(r^i(v)/c)}{\exp(r(v)/c)}\right) \nonumber \\
    &= \Pr[\bar{v}=v\mid\hat{c},\hat{d}]\left(1-\exp\left(\frac{-(r(v)-r^i(v))}{c}\right)\right) \nonumber \\
    &\leq \Pr[\bar{v}=v\mid\hat{c},\hat{d}]\cdot \frac{r(v)-r^i(v)}{c},\label{eq:exptolinear}
\end{align}
where the first equality is from the algorithm and the inequality is from $1-\exp(-x)\leq x$.
Therefore, we have
\begin{align*}
    \frac{1}{n}\sum_{i=1}^{n}\sum_{v\in V_{d}\setminus S_i}\frac{\exp(r(v)/c)-\exp(r^i(v)/c)}{\sum_{v'\in V_{d}}\exp(r(v')/c)}
    &\leq \frac{1}{n}\sum_{v\in V_{d}}\Pr[\bar{v}=v\mid\hat{c},\hat{d}]\sum_{i\colon S_i\not \ni v} \frac{r(v)-r^i(v)}{c}\\
    &\leq \frac{1}{n}\sum_{v\in V_{d}}\Pr[\bar{v}=v\mid\hat{c},\hat{d}]\cdot \frac{Kr(v)}{c}\\
    &\leq \frac{1}{n}\cdot \frac{K}{c} \cdot \OPT(V)\\
    &\leq \frac{1}{n}\cdot K\epsilon^{-1}\log\left(|V|\epsilon^{-1}\right),
\end{align*}
where the first inequality is from~\eqref{eq:exptolinear}, the second inequality is from Lemma~\ref{lem:sumdiff}, the third inequality is from $r(v)\leq \OPT(V)$ and the last inequality is from the definition of $c$.
\end{proof}

The next lemma bounds the second term of~\eqref{eq:pseudotriangle}.
\begin{lemma}\label{lem:triangle_2}
Conditioned on having chosen $c=c^i=\hat{c}$, we have
\[
    \frac{4}{|V|}\int_{\frac{1}{2}|V|}^{\frac{3}{4}|V|}\left(\frac{1}{n}\sum_{i=1}^{n}\sum_{v\in V_d\setminus S_i}\max\left(0,\frac{\exp(r^i(v)/c)}{\sum_{v'\in V_{d}}\exp(r(v')/c)}-\frac{\exp(r^i(v)/c)}{\sum_{v'\in V_{d}^i}\exp(r^i(v')/c)}\right)\right)\mathrm{d}d
    \leq \frac{1}{n}\cdot 4K.
\]
\end{lemma}
\begin{proof}
First, we have
\begin{align}
    &\max\left(0,\frac{\exp(r^i(v)/c)}{\sum_{v'\in V_{d}}\exp(r(v')/c)}-\frac{\exp(r^i(v)/c)}{\sum_{v'\in V_{d}^i}\exp(r^i(v')/c)}\right) \nonumber \\
    &\leq \max\left(0,\frac{\exp(r^i(v)/c)}{\sum_{v'\in V_{d}}\exp(r(v')/c)}-\frac{\exp(r^i(v)/c)}{\sum_{v'\in V_{d+|S_i|}}\exp(r^i(v')/c)}\right) \nonumber \\
    &\leq \max\left(0,\frac{\exp(r^i(v)/c)}{\sum_{v'\in V_{d}}\exp(r(v')/c)}-\frac{\exp(r^i(v)/c)}{\sum_{v'\in V_{d+|S_i|}}\exp(r(v')/c)}\right) \nonumber \\
    &= \frac{\exp(r^i(v)/c)}{\sum_{v'\in V_{d}}\exp(r(v')/c)}-\frac{\exp(r^i(v)/c)}{\sum_{v'\in V_{d+|S_i|}}\exp(r(v')/c)} \nonumber \\
    &\leq \frac{\exp(r(v)/c)}{\sum_{v'\in V_{d}}\exp(r(v')/c)}-\frac{\exp(r(v)/c)}{\sum_{v'\in V_{d+|S_i|}}\exp(r(v')/c)},\label{eq:errord1}
\end{align}
where the first inequality is from $V^i_{d}\subseteq V_{d+|S_i|}$ that is from $\max\left(|V^i_{-v}|,|V^i_{+v}|\right)\geq \max\left(|V_{-v}|,|V_{+v}|\right)+|S_i|$, the second and the last inequality is from $r^i(v)\leq r(v)$, and the equality is from $V_d\subseteq V_{d+|S_i|}$.

By taking the expectation over $d$, we have
\begin{align*}
    &\frac{4}{|V|}\int_{\frac{1}{2}|V|}^{\frac{3}{4}|V|}\left(\frac{1}{n}\sum_{i=1}^{n}\sum_{v\in V_d\setminus S_i}\max\left(0,\frac{\exp(r(v)/c)}{\sum_{v'\in V_{d}}\exp(r(v')/c)}-\frac{\exp(r^i(v)/c)}{\sum_{v'\in V_{d}^i}\exp(r(v')/c)}\right)\right)\mathrm{d}d\\
    &\leq \frac{4}{|V|}\int_{\frac{1}{2}|V|}^{\frac{3}{4}|V|}\left(\frac{1}{n}\sum_{i=1}^{n}\sum_{v\in V_d\setminus S_i}\left(\frac{\exp(r(v)/c)}{\sum_{v'\in V_{d}}\exp(r(v')/c)}-\frac{\exp(r(v)/c)}{\sum_{v'\in V_{d+|S_i|}}\exp(r(v')/c)}\right)\right)\mathrm{d}d\\
    &=\frac{1}{n}\sum_{i=1}^{n}\left(\frac{4}{|V|}\int_{\frac{1}{2}|V|}^{\frac{3}{4}|V|}\sum_{v\in V_d\setminus S_i}\frac{\exp(r(v)/c)}{\sum_{v'\in V_d}\exp(r(v')/c)}\mathrm{d}d-\frac{4}{|V|}\int_{\frac{1}{2}|V|+|S_i|}^{\frac{3}{4}|V|+|S_i|}\sum_{v\in V_d\setminus S_i}\frac{\exp(r(v)/c)}{\sum_{v'\in V_{d}}\exp(r(v')/c)}\mathrm{d}d\right)\\
    &=\frac{1}{n}\sum_{i=1}^{n}\left(\frac{4}{|V|}\int_{\frac{1}{2}|V|}^{\frac{1}{2}|V|+|S_i|}\sum_{v\in V_d\setminus S_i}\frac{\exp(r(v)/c)}{\sum_{v'\in V_d}\exp(r(v')/c)}\mathrm{d}d-\frac{4}{|V|}\int_{\frac{3}{4}|V|}^{\frac{3}{4}|V|+|S_i|}\sum_{v\in V_d\setminus S_i}\frac{\exp(r(v)/c)}{\sum_{v'\in V_{d}}\exp(r(v')/c)}\mathrm{d}d\right)\\
    &\leq \frac{1}{n}\sum_{i=1}^{n}\left(\frac{4}{|V|}\int_{\frac{1}{2}|V|}^{\frac{1}{2}|V|+|S_i|}\sum_{v\in V_d\setminus S_i}\frac{\exp(r(v)/c)}{\sum_{v'\in V_d}\exp(r(v')/c)}\mathrm{d}d\right)\\
    &\leq \frac{1}{n}\sum_{i=1}^{n}\frac{4}{|V|}\cdot |S_i| \leq \frac{1}{n}\cdot \frac{4}{|V|}\cdot K|V|=\frac{1}{n}\cdot 4K,
\end{align*}
where the first inequality is from~\eqref{eq:errord1}, the second equality is obtained by cancelling the integral intervals, third inequality is from
\[
    \sum_{v\in V_d\setminus S_i}\frac{\exp(r(v)/c)}{\sum_{v'\in V_d}\exp(r(v')/c)}\leq \frac{\sum_{v\in V_d\setminus S_i}\exp(r(v)/c)}{\sum_{v'\in V_d}\exp(r(v')/c)}\leq 1,
\]
and the fourth inequality is from $\sum_{i=1}^n|S_i|\leq K|V|$.
\end{proof}

Combining~\eqref{eq:pseudotriangle} and Lemmas~\ref{lem:triangle_1} and~\ref{lem:triangle_2} yields the following.
\begin{lemma}\label{lem:deteletriangle}
Assume Algorithm~\ref{alg:main} chose the parameter $c=c^i=\hat{c}$. Then, we have
\[
    \frac{4}{|V|}\int_{\frac{1}{2}|V|}^{\frac{3}{4}|V|}\left(\frac{1}{n}\sum_{i=1}^{n}\sum_{v\in V_{\hat{d}}\setminus S_i}\max\left(0,\Pr[\bar{v}=v\mid\hat{c},\hat{d}]-\Pr[\bar{v}^i=v\mid\hat{c},\hat{d}]\right)\right)\mathrm{d}\hat{d}\leq \frac{1}{n}\cdot 2K\epsilon^{-1}\log\left(|V|\epsilon^{-1}\right).
\]
\end{lemma}
\begin{proof}
By applying Lemmas~\ref{lem:triangle_1} and~\ref{lem:triangle_2} on~\eqref{eq:pseudotriangle}, we have
\begin{align*}
    &\frac{4}{|V|}\int_{\frac{1}{2}|V|}^{\frac{3}{4}|V|}\left(\frac{1}{n}\sum_{i=1}^{n}\sum_{v\in V_{\hat{d}}\setminus S_i}\max\left(0,\Pr[\bar{v}=v\mid\hat{c},\hat{d}]-\Pr[\bar{v}^i=v\mid\hat{c},\hat{d}]\right)\right)\mathrm{d}\hat{d}\\
    &\leq \frac{4}{|V|}\int_{\frac{1}{2}|V|}^{\frac{3}{4}|V|}\left(\frac{1}{n}\sum_{i=1}^{n}\sum_{v\in V_{\hat{d}}\setminus S_i}\left(\frac{\exp(r(v)/c)-\exp(r^i(v)/c)}{\sum_{v'\in V_{\hat{d}}}\exp(r(v')/c)}\right.\right.\\
    &\quad \quad \left.\left.+\max\left(0,\frac{\exp(r^i(v)/c)}{\sum_{v'\in V_{\hat{d}}}\exp(r(v')/c)}-\frac{\exp(r^i(v)/c)}{\sum_{v'\in V_{\hat{d}}^i}\exp(r^i(v')/c)}\right)\right)\right)\mathrm{d}\hat{d}\\
    &\leq \frac{4}{|V|}\int_{\frac{1}{2}|V|}^{\frac{3}{4}|V|}\frac{1}{n}K\epsilon^{-1}\log\left(|V|\epsilon^{-1}\right)\mathrm{d}\hat{d}+\frac{1}{n}\cdot 4K\\
    &= \frac{1}{n}\left(K\epsilon^{-1}\log\left(|V|\epsilon^{-1}\right)+4K\right)\leq \frac{1}{n}\cdot 2K\epsilon^{-1}\log\left(|V|\epsilon^{-1}\right),
\end{align*}
where the first inequality is from~\eqref{eq:pseudotriangle}, the second inequality is from Lemma~\ref{lem:triangle_1} and Lemma~\ref{lem:triangle_2}, and the last inequality is from $\epsilon<0.2$.
\end{proof}

Now we complete the analysis by combining all the aforementioned lemmas.

\begin{proof}[Proof of Lemma~\ref{lem:pivottvd}]
We have
\begin{align*}
    &\frac{1}{n}\sum_{i=1}^{n}\TV(\bar{v},\bar{v}^i)\\
    &\leq \sup_{\hat{c}\in [B,2B]}\left(\frac{4}{|V|}\int_{\frac{1}{2}|V|}^{\frac{3}{4}|V|}\left(\frac{1}{n}\sum_{i=1}^{n}\TV\left((\bar{v} \mid c = \hat{c},d=\hat{d}),(\bar{v}^i \mid c^i = \hat{c},d^i=\hat{d})\right)\right)\mathrm{d}\hat{d}\right)+\frac{1}{n}\cdot 9K\log\left(|V|\epsilon^{-1}\right)\\
    &= \sup_{\hat{c}\in [B,2B]}\left(\frac{4}{|V|}\int_{\frac{1}{2}|V|}^{\frac{3}{4}|V|}\left(\frac{1}{n}\sum_{i=1}^{n}\sum_{v\in S_i}\Pr[\bar{v}=v\mid\hat{c},\hat{d}]\right.\right.\\
    &\quad \quad \left.\left.+\frac{1}{n}\sum_{i=1}^{n}\sum_{v\in V_{\hat{d}}\setminus S_i}\max\left(0,\Pr[\bar{v}=v\mid\hat{c},\hat{d}]-\Pr[\bar{v}^i=v\mid\hat{c},\hat{d}]\right)\right)\mathrm{d}\hat{d}\right)
    +\frac{1}{n}\cdot 9K\log\left(|V|\epsilon^{-1}\right)\\
    &\leq \sup_{\hat{c}\in [B,2B]}\left(\frac{4}{|V|}\int_{\frac{1}{2}|V|}^{\frac{3}{4}|V|}\frac{1}{n}\cdot K\mathrm{d}\hat{d}+\frac{1}{n}\cdot 2K\epsilon^{-1}\log\left(|V|\epsilon^{-1}\right)\right)+\frac{1}{n}\cdot 9K\log\left(|V|\epsilon^{-1}\right)\\
    &= \frac{1}{n}\left(K+2K\epsilon^{-1}\log\left(|V|\epsilon^{-1}\right)+9K\log\left(|V|\epsilon^{-1}\right)\right)\\
    &\leq \frac{1}{n}\cdot 3K\epsilon^{-1}\log\left(|V|\epsilon^{-1}\right),
\end{align*}
where the first inequality is Lemma~\ref{lem:impact_cd}, the first inequality is from~\eqref{eq:sumprob}, the second inequality is from Lemma~\ref{lem:deletev} and Lemma~\ref{lem:deteletriangle}, and the last inequality is from $\epsilon\leq 0.2$.
\end{proof}


\section{Direct Applications}\label{sec:application}

In this section, we provide stable-on-average algorithms for several DP problems by reducing them to the maximum chain problem. 
For each of the problems discussed here, we construct a vertex-weighted directed graph $G=(V,E,w)$ and antichains $S_1,\ldots,S_n$ from the instance $A$ of the original problem.
If they satisfy the following conditions, then for any $\delta>0$, we automatically obtain a stable-on-average polynomial-time $(1-\delta)$-approximation algorithm by Theorem~\ref{thm:main}:
\begin{itemize}
    \itemsep=0pt
    \item $G$ is acyclic and transitive.
    \item There is a surjective map from the chains in $G$ to the solutions for the original instance preserving the weight.
    \item For each element $i \in A$ in the original instance, the subgraph of $G$ induced by $V\setminus S_i$ is isomorphic to the graph $G^i$ constructed from the instance $A\setminus \{i\}$.
    \item There exists a constant $K > 0$ such that every vertex $v\in V$ belongs to at least one and at most $K$ of $S_i$'s.
\end{itemize}
The resulting average sensitivity is $O(K\delta^{-1}\log^3 |V|)$.
In the second condition, the surjectivety is necessary to ensure that any solution for the original instance is represented by some chain in $G$ and hence that the optimal values of the two problems are equal.


\subsection{Longest Increasing Subsequence}\label{sec:LIS}

The longest increasing subsequence problem defined below showcases our methodology for obtaining stable-on-average algorithms.
\begin{prb}[Longest Increasing Subsequence]
Let $A=(a_1,\dots, a_n)$ be a sequence of integers.
Find the largest set $X = \{i_1,\dots, i_t\}$ of indices such that $i_1<\cdots<i_t$ and $a_{i_1}<\cdots<a_{i_t}$.
\end{prb}
The average sensitivity of an algorithm $\Call{Alg}{}$ for the longest increasing subsequence problem is defined as $\frac{1}{n}\sum_{i=1}^n d_{\mathrm{EM}}(\Call{Alg}{A}, \Call{Alg}{A^i})$, where for $i \in \{1,2,\ldots,n\}$, $A^i = (a_1,\dots, a_{i-1},a_{i+1},\dots a_n)$ is the sequence obtained from $A$ by dropping $a_i$.


We construct the graph $G=(V,E,w)$ to apply Algorithm~\ref{alg:main} on as
\begin{align*}
    V=\{1,\dots, n\},
    \quad
    E=\{(i,j)\colon i<j, a_i<a_j\},
    \quad
    w(i) = 1 \text{ for any } i \in V.\;
\end{align*}
The graph $G$ represents a textbook DP shown in Algorithm~\ref{alg:LIS}.
At Lines~\ref{line:lis-argmax} and~\ref{line:lis-argmax2} of Algorithm~\ref{alg:LIS}, we use the convention that for a condition $P$, $\DP[\mathrm{argmax}_{i:\text{$i$ satisfies $P$}}|\DP[i]|]$ (or $\DP[i^*]$ with $i^* \leftarrow \mathrm{argmax}_{i:\text{$i$ satisfies $P$}}|\DP[i]|$) denotes the empty set if no $i$ satisfies $P$.
We use the same convention in the rest of this section.


\begin{algorithm}[t!]
\caption{Textbook algorithm for the longest increasing subsequence problem}\label{alg:LIS}
\Procedure{\emph{\Call{LIS}{$A$}}}{
    $\DP[i]\leftarrow \emptyset$ for all $i=1,\dots, n$\;
    \For{$j=1,\dots, n$}{
        $i^*\leftarrow \mathrm{argmax}_{1 \leq i < j: a_i < a_j}|\DP[i]|$\;
        $\DP[j]\leftarrow \DP[i^*]\cup \{j\}$\;\label{line:lis-argmax}
    }
    \Return $\DP\left[\mathrm{argmax}_{1 \leq i \leq n}|\DP[i]|\right]$\;\label{line:lis-argmax2}
}
\end{algorithm}

We now check the four conditions required to apply Theorem~\ref{thm:main}.
Clearly $G$ is acyclic and transitive.
A chain $(i_1,\dots, i_t)$ in $G$ is bijectively mapped to a feasible solution $\{i_1,\dots, i_t\}$ for the original instance.
Since $E$ is defined solely by the inequality relation over integers, the subgraph of $G$ induced by $V \setminus \{i\}$ is isomorphic to the graph constructed from the instance $A^i$.
By applying Theorem~\ref{thm:main} on $G$, $S_i=\{i\}$ for $i=1,\dots, n$, and $K=1$, we obtain the following:
\begin{corollary}
For any $\delta>0$, there is a polynomial-time $(1-\delta)$-approximation algorithm for the longest increasing subsequence problem with average sensitivity $O(\delta^{-1}\log^3 n)$.
\end{corollary}

\subsection{Interval Scheduling}\label{sec:intervalscheduling}
The interval scheduling problem, defined below, can be used to model scheduling tasks.
\begin{prb}[Interval Scheduling]
Let $A=\{[l_1,r_1),[l_2,r_2),\ldots,[l_n,r_n)\}$ be a set of nonempty intervals over $\mathbb{R}$ and let $w(1),\ldots,w(n)$ be positive weights.
Find a subset $X \subseteq \{1,2,\ldots,n\}$ of indices that maximizes the total weight $\sum_{i\in X}w(i)$ such that for any distinct $i,j\in X$, the intervals $[l_i,r_i)$  and $[l_j,r_j)$ are disjoint.
\end{prb}
The average sensitivity of an algorithm $\Call{Alg}{}$ for the interval scheduling problem is defined as $\frac{1}{n}\sum_{i=1}^n d_{\mathrm{EM}}(\Call{Alg}{A}, \Call{Alg}{A^i})$, where for $i \in \{1,2,\ldots,n\}$, $A^i$ is the set of intervals obtained from $A$ by dropping the $i$-th interval $[l_i,r_i)$.


We construct the directed graph $G=(V,E,w)$ to apply Algorithm~\ref{alg:main} on as
\begin{align*}
    V=\{1,\dots, n\}, \quad
    E=\{(i,j)\colon r_i\leq l_j\},
\end{align*}
and $w$ is the same weight function as the one for the original instance.
The graph $G$ represents a textbook DP shown in Algorithm~\ref{alg:IntervalScheduling}.

\begin{algorithm}[t!]
\caption{Textbook algorithm for the interval scheduling problem}\label{alg:IntervalScheduling}
\Procedure{\emph{\Call{IntervalScheduling}{$A,w$}}}{
    $\DP[i]\leftarrow \emptyset$ for all $i=1,\dots, n$\;
    Sort input intervals in ascending order of $r_i$ so that $r_1\leq \cdots \leq r_n$\;
    \For{$j=1,\dots, n$}{
        $i^*\leftarrow \mathrm{argmax}_{i:r_i \leq l_j}w(\DP[i])$\;
        $\DP[j]\leftarrow \DP[i^*]\cup \{[l_j,r_j)\}$\;
    }
    \Return $\DP\left[\mathrm{argmax}_{1 \leq i \leq n} w(\DP[i])\right]$\;
}
\end{algorithm}

We now check the four conditions required to apply Theorem~\ref{thm:main}.
Clearly, $G$ is acyclic and transitive.
A chain $(i_1,\dots, i_t)$ in $G$ is bijectively mapped to a solution $\{i_1,\dots, i_t\}$ for the original instance.
Since $E(G)$ is defined only by the inequality relation over integers, the subgraph of $G$ induced by $V\setminus \{i\}$ is isomorphic to the graph constructed from the instance $A^i$ (and $w$ restricted to $\{1,2,\ldots,n\} \setminus \{i\}$).
By applying Theorem~\ref{thm:main} on $G$, $S_i=\{i\}$ for $i=1,\dots, n$, and $K=1$, we obtain the following.
\begin{corollary}
For any $\delta>0$, there exists a polynomial-time $(1-\delta)$-approximation algorithm for the interval scheduling problem with average sensitivity $O(\delta^{-1}\log^3 n)$.
\end{corollary}

\subsection{Longest Common Subsequence}\label{sec:LCS}
The longest common subsequence problem is defined as follows:
\begin{prb}[Longest Common Subsequence~\cite{algorithmintroduction}]
Let $A_1,\dots, A_k$ be strings over some alphabet.
Find a set  $X = \{(p_{1,1},\dots,p_{1,k}),\dots, (p_{t,1},\dots, p_{t,k})\}$ of index lists of the same length that maximize $t$ subject to $1\leq p_{i,1} < \cdots < p_{i,t} \leq |A_i|$ for all $i$ and $A_{1,p_{1,j}}=\cdots=A_{k,p_{k,j}}$ for all $j$.
\end{prb}

The distance between two solutions $X_1, X_2$ is defined by $|X_1\triangle X_2|$, where we regard $X_i$ as a set of lists, each consisting of $k$ indices, and two lists are regarded as equal if they consist of the same set of elements in the same order.
Note that this distance upper-bounds the edit distance~\cite{dan1997algorithms} of the two outputs regarded as strings.

We consider the situation that one of the $|A_1|+\cdots+|A_k|$ letters in the input is deleted.
For $i\in\{1,2,\ldots,k\}$ and $j \in \{1,2,\ldots,|A_i|\}$, let $A_i^j$ denote the string $A_{i,1}\dots A_{i,j-1}A_{i,j+1}\dots, A_{i,|A_i|}$
Then, the average sensitivity of an algorithm $\Call{Alg}{}$ is defined as
\begin{align*}
    \frac{1}{\sum_{i=1}^k |A_i|}\sum_{i=1}^{k}\sum_{j=1}^{|A_i|}\EM\left(\Call{Alg}{A_1,\dots,A_k},\Call{Alg}{A_1,\dots, A_{i-1},A_i^j,A_{i+1},\dots, A_{k}}\right).
\end{align*}
We construct the graph $G=(V,E,w)$ to apply Algorithm~\ref{alg:main} on as
\begin{align*}
    V&=\{(p_1,\ldots, p_k)\colon p_i\in \{1,\dots, |A_i|\}\text{ for all $i=1,\dots, k$}, A_{1p_1}=\cdots=A_{kp_k}\}\},\\
    E&=\{((p_1,\ldots, p_k),(q_1,\dots, q_k))\colon p_1 < q_1,\ldots, p_k < q_k\}, \\
    w(v) & = 1 \text{ for every } v \in V.
\end{align*}
The graph $G$ represents a DP due to Wagner and Fischer~\cite{LCS_wagner} shown in Algorithm~\ref{alg:LCS}.
\begin{algorithm}[t!]
\caption{Wagner and Fischer's algorithm}\label{alg:LCS}
\Procedure{\emph{\Call{LCS}{$A_1,\dots,A_k$}}}{
    $\DP[p_1,\dots, p_k]\leftarrow \emptyset$ for all $1\leq p_1\leq |A_1|,\dots, 1\leq p_k\leq |A_k|$\;
    \For{$q_1=1,\dots, |A_1|$}{
        $\ddots$\\
        \For{$q_k=1,\dots, |A_k|$}{
            \If{$A_{1,q_1}=\cdots=A_{k,q_k}$}{
                $(p_1^*,\dots, p_k^*)\leftarrow \mathrm{argmax}_{p_1<q_1,\dots, p_k<q_k, A_{1,p_1}=\cdots=A_{k,p_k}}|\DP[p_1,\dots, p_k]|$\;
                $\DP[q_1,\dots, q_k]\leftarrow \DP[p_1^*,\dots, p_k^*]\cup \{q_1,\dots, q_k\}$\;
            }
        }
    }
    \Return $\DP\left[\mathrm{argmax}_{A_{1,p_1}=\cdots=A_{k,p_k}}|\DP[p_1,\dots, p_k]|\right]$\;
}
\end{algorithm}

We now check the four conditions required to apply Theorem~\ref{thm:main}.
Clearly, $G$ is acyclic and transitive.
A chain $(p_{1,1},\dots, p_{1,k}),\dots, (p_{t,1},\dots, p_{t,k})$ in $G$ is bijectively mapped to a solution $\{(p_{1,1},\dots, p_{1,k}),\dots, (p_{t,1},\dots, p_{t,k})\}$ for the original instance.
For $i=1,\dots, k$ and $j=1,\dots, |A_i|$, let
\begin{align}
S_{i,j}=\{(p_1,\dots, p_k)\in V\colon p_i=j\}.
\end{align}
Note that $S_{i,j}$ is an antichain in $G$.
Since $E$ is defined only by the inequality relation over integers, the subgraph of $G$ induced by $V\setminus S_{i,j}$ is isomorphic to the graph constructed from the instance $(A_1,\ldots,A_{i-1},A_i^j,A_{i+1},\ldots,A_k)$.
By applying Theorem~\ref{thm:main} on $G$, $S_{i,j}$ for all $i=1,\dots, k$ and $j=1,\dots, |A_i|$, and $K=k$, we obtain the following.
\begin{corollary}
For any $\delta>0$, there is a polynomial-time $(1-\delta)$-approximation algorithm for the longest common subsequence problem with with average sensitivity $O\left(k\delta^{-1}\log^3 \left(\prod_{i=1}^{k}|A_i|\right)\right)$.
\end{corollary}


\subsection{Longest Palindromic Subsequence}\label{sec:LPS}
The \emph{longest palindromic subsequence problem}~\cite{palindrome_chowdhury,algorithmintroduction} searches for a longest palindrome that is a subsequence of the input string, where a string is called a \emph{palindrome} if it is identical to itself reversed.
The problem is formally defined as follows:
\begin{prb}[Longest Palindromic Subsequence~\cite{algorithmintroduction}]
Let $A$ be a string of length $n$ over some alphabet.
Find a largest set of indices $\{i_1,\dots, i_t\}$ such that $i_1<\cdots<i_t$ and the substring $A_{i_1}\ldots A_{i_t}$ is a palindrome.
\end{prb}
The average sensitivity of an algorithm $\Call{Alg}{}$ for the longest palindromic subsequence problem is defined as $\frac{1}{n}\sum_{i=1}^n d_{\mathrm{EM}}(\Call{Alg}{A}, \Call{Alg}{A^i})$, where for $i \in \{1,2,\ldots,n\}$, $A^i$ is the substring obtained from $A$ by dropping the $i$-th letter.


We construct the graph $G=(V,E,w)$ to apply Algorithm~\ref{alg:main} on as
\begin{align*}
    V(G)&=\{(p,q)\colon 1\leq p\leq q\leq n, A_p=A_q\},\\
    E(G)&=\{((p_1,q_1),(p_2,q_2))\colon p_1<p_2\leq q_2<q_1\},\\
    w((p,q)) & = \begin{cases}
    2 & \text{if } p < q,\\
    1 & \text{if } p = q.
    \end{cases}
\end{align*}
The graph $G$ represents the folklore DP shown in Algorithm~\ref{alg:LPS}, which is given as an exercise in~\cite{algorithmintroduction}.
\begin{algorithm}[t!]
\caption{Folklore algorithm for the longest palindromic subsequence problem}\label{alg:LPS}
\Procedure{\emph{\Call{LongestPalindromicSubsequence}{$A$}}}{
    $\DP[p][q]\leftarrow \emptyset$ for all $1\leq p\leq q\leq n$\;
    \For{$p_2=1,\dots, n$}{
        $\DP[p_2][p_2]=\{p_2\}$\;
        \For{$q_2=p_2+1,\dots, n$}{
            \If{$A_{p_2}=A_{q_2}$}{
                $(p_1^*,q_1^*)\leftarrow \mathrm{argmax}_{p_1<p_2\leq q_2<q_1, A_{p_1}=A_{q_1}}\DP[p_1][q_1]$\;
                $\DP[p_2][q_2]\leftarrow \DP[p_1^*][q_1^*]\cup \{p_2,q_2\}$\;
            }
        }
    }
    \Return $\DP\left[\mathrm{argmax}_{p\leq q, A_{p}=A_{q}}|\DP[p][q]|\right]$\;
}
\end{algorithm}

We now check the four conditions required to apply Theorem~\ref{thm:main}.
Clearly, $G$ is acyclic and transitive.
A chain $((p_1,q_1),\dots,(p_t,q_t))$ in $G$ is bijectively mapped to a solution $(p_1,\dots, p_t, q_t,\dots, q_1)$ for the original string if $p_t < q_t$ and $(p_1,\dots, p_t=q_t,\dots, q_1)$ if $(p_t=q_t)$.
For $i=1,\dots, n$, let
\begin{align}
S_i=\{(p,q)\in V\colon p=i\text{ or }q=i\}.
\end{align}
Note that $S_i$ is an antichain in $G$.
Since $E$ is defined only by the inequality relation over integers, the graph induced by $V\setminus S_i$ is isomorphic to the graph constructed from the string $A^i$.
By applying Theorem~\ref{thm:main} on $G$, $S_i$ for $i=1,\dots, n$, and $K=2$, we have the following:
\begin{corollary}
For any $\delta>0$, there is a polynomial-time $(1-\delta)$-approximation algorithm for the longest palindromic subsequence problem with average sensitivity $O(\delta^{-1}\log^3 n)$.
\end{corollary}

\subsection{Knapsack Problem with Integer Cost}\label{sec:knapsack}
The knapsack problem~\cite{kellerer2004knapsack} is one of the most classical optimization problems and is defined as follows.
\begin{prb}[Knapsack Problem~\cite{kellerer2004knapsack}]
Let $A$ be a set of $n$ items and $C$ be a cost limit. The items are numbered $1,\dots, n$. Each item $i$ has a cost $c(i)$ and a weight $w(i)$.
Find a subset $X$ of $A$ that maximizes the total weight $\sum_{i\in X}w(i)$ subject to $\sum_{i\in X}c(i)\leq C$.
\end{prb}
As with other problems discussed in this section, the average sensitivity of an algorithm $\Call{Alg}{}$ for the knapsack problem is defined as $\frac{1}{n}\sum_{i=1}^n d_{\mathrm{EM}}(\Call{Alg}{A}, \Call{Alg}{A^i})$, where for $i \in \{1,2,\ldots,n\}$, $A^i$ is the instance obtained from $A$ by deleting the $i$-th item.

Here, we consider a special case of the knapsack problem with an additional constraint that $C$ and $c(i)$ are integers.
The graph $G=(V,E,w)$ to apply Algorithm~\ref{alg:main} on is defined by
\begin{align*}
    V&=\{(i,p)\colon i\in \{1,\dots, n\}, p\in \{c(i),\dots, C\}\},\\
    E&=\{((i_1,p_1),(i_2,p_2))\colon i_1<i_2, p_1+c(i_2)\leq p_2\},\\
    w(i,p)&=w(i) \text{ for any } (i,p) \in V.
\end{align*}
The graph $G$ represents a textbook algorithm shown in Algorithm~\ref{alg:knapsack}.
\begin{algorithm}[t!]
\caption{Textbook algorithm for the knapsack problem}\label{alg:knapsack}
\Procedure{\emph{\Call{Knapsack}{$A$}}}{
    $\DP[i][p]\leftarrow \emptyset$ for all $i=1,\dots, n$ and $p=0,\dots, C$\;
    \For{$i_2=1,\dots, n$}{
        \For{$p_2=c(i_2)\dots, C$}{
            $(i_1^*,p_1^*)\leftarrow \mathrm{argmax}_{i_1<i_2,p_1+c(i_2)<p_2}w(\DP[i_1][p_1])$\;
            $\DP[i_2][p_2]\leftarrow \DP[i_1^*][p_1^*]\cup \{i_2\}$\;
        }
    }
    \Return $\DP\left[\mathrm{argmax}_{i=1}^{n}w(\DP[i][C])\right]$\;
}
\end{algorithm}

We now check the four conditions required to apply Theorem~\ref{thm:main}.
Clearly, $G$ is acyclic and transitive.
A chain $((i_1,p_1),\dots, (i_t,p_t))$ is bijectively mapped to a solution $\{i_1,\dots, i_t\}$ for the original instance.
For $i=1,\ldots, n$, let
\begin{align}
S_i=\{(i',p)\in V\colon i=i'\}.
\end{align}
Note that $S_i$ is an antichain in $G$.
Since $E$ is defined only by the inequality relation over integers, the subgraph induced by $V\setminus S_i$ is isomorphic to the graph constructed from the instance $A^i$.
By applying Theorem~\ref{thm:main} on $G$, $S_i$ for $i=1,\dots, n$, and $K=1$, we obtain the following.
\begin{corollary}
For any $\delta>0$, there is a polynomial-time $(1-\delta)$-approximation algorithm for the knapsack problem with average sensitivity $O(\delta^{-1}\log^3 (nC))$.
\end{corollary}


\section{RNA Folding}\label{sec:RNA}
In this section, we provide a stable-on-average algorithm for the RNA folding problem.



\begin{theorem}\label{thm:rna-folding}
For any $\delta>0$, there exists a quasi-polynomial-time $(1-\delta)$-approximation algorithm for the RNA folding problem with average sensitivity $O(\delta^{-1}\log^7 n)$.
\end{theorem}




The stable-on-average algorithm presented here is far more complicated than the algorithms in the previous sections.
First, we explain why a naive algorithm does not work.
Algorithm~\ref{alg:naive} is the algorithm for computing the optimal value of the RNA folding problem according to Nussinov and Jacobson~\cite{RNA_nussinov}. Here, $\DP[i][j]$ is supposed to store the optimal value for the substring $A_i A_{i+1}\cdots A_j$.
Since we should calculate the maximum over the sum of two values stored in $\DP$ at Line~\ref{line:2d1ddp} in Algorithm~\ref{alg:naive}, the DP performed in Algorithm~\ref{alg:naive} cannot be (easily) written as MWC\@.
To resolve this issue, we convert this DP into another DP that can be regarded as a maximum chain problem on DAG\@.


\begin{algorithm}[t!]
\caption{Nussinov and Jacobson's algorithm~\cite{RNA_nussinov}}\label{alg:naive}
\Procedure{\emph{\Call{CubicRNAFolding}{$A$}}}{
    $\DP[i][j]\leftarrow 0$ for all $1\leq i,j\leq n$\;
    \For{$d=2,\dots,n$}{
        \For{$i=1,\dots, n-d+1$}{
            $j\leftarrow i+d-1$\;
            $\DP[i][j]\leftarrow \max_{i\leq k<j}(\DP[i][k]+\DP[k+1][j])$\;\label{line:2d1ddp}
            \If{$\left(A_i,A_j\right)\in \mathcal{R}$}{
                $\DP[i][j]\leftarrow \max(\DP[i][j],\DP[i+1][j-1]+1)$\;
            }
        }
    }
    \Return $\DP[1][n]$\;
}
\end{algorithm}

\subsection{Graph Construction}

Consider constructing a graph $G=(V,E,w)$ such that a chain in $G$ can surjectively be mapped to a solution for the original string with the same weight.
The simplest (unsuccessful) idea for constructing $G$ is to introduce a vertex for each pair $(l,r)$ with $(A_l,A_r)\in \mathcal{R}$ and define $E$ so that we can recover a solution for the original string from a chain in $G$, e.g.,
\begin{align*}
    V&=\{(l,r)\colon 1\leq l<r\leq n, (A_l,A_r)\in \mathcal{R}\},\\
    E&=\{((l_1,r_1),(l_2,r_2))\colon r_1<l_2 \text{ or } l_1<l_2<r_2<r_1\}.
\end{align*}
However, this idea does not work because a chain in $G$ may not correspond to a feasible solution.
For example, if $A=\text{abbcac}$, then there should be edges $((1,5),(2,3))$ and $((2,3),(4,6))$ in $G$ but $((1,5),(4,6))$ should not exist in $G$ because $((1,5),(4,6))$ is a pseudoknot. However $G$ is no longer transitive.

We can think of another (unsuccessful) idea for constructing $G$ that resolves the aforementioned issue.
Here, a vertex in $G$ corresponds to a list of pairs $((l_1,r_1),\ldots,(l_k,r_k))$ with $(A_{l_i},A_{r_i}) \in \mathcal{R}$ for every $i$ such that $l_i < l_j < r_j < l_i$ for every $i < j$, which encodes the peeling structure of a solution.
To define the edge set of $G$, we define a partial order $\prec'$ over index pairs such that $(l,r) \prec' (l',r')$ holds if $r<l'$.
Then, we define $G=(V,E,w)$ as
\begin{align*}
    V&=\{((l_1,r_1),\ldots,(l_k,r_k))\colon (A_{l_j},A_{r_j})\in \mathcal{R} \text{ for all }j=1,\dots, k, l_1<\cdots<l_k<r_k<\cdots<r_1\},\\
    E&=\{(((l_1,r_1),\ldots,(l_k,r_k)),((l'_1,r'_1),\dots,(l'_k,r'_k)))\colon \\
    &\quad \quad \text{the former is lexicographically smaller than the latter under the partial order $\prec'$}\},
\end{align*}
where for two lists $v=(x_1,\dots, x_k), v'=(x'_1,\dots, x'_{k'})$ of pairs, $v$ is lexicographically smaller than $v'$ if $v$ is a prefix of $v'$ or $x_j\prec' x'_j$ holds for the first index $j$ with $x_j\neq x'_j$.

In this example, a chain $((l_{1,1},r_{1,1}),\dots, (l_{1,k_1},r_{1,k_1})),\dots, ((l_{t,1},r_{t,1}),\dots, (l_{t,k_t},r_{t,k_t}))$ of $G$ can be surjectively mapped to a solution $\{(l_{1,k_1},r_{1,k_1}),\dots, (l_{t,k_t},\dots, r_{t,k_t})\}$ of the original problem.
Furthermore, if we set the weight of all vertices in $G$ as $1$, this mapping preserves the weight.
However, an issue of this reduction is that $|V|$ can be exponential in $n$.
Since the average sensitivity of $\Call{MWC}{}$ is polylogarithm in $|V|$, the average sensitivity on $G$ obtained by applying Theorem~\ref{thm:main} becomes polynomial of $n$, which exceeds the trivial bound of $n$.

However, we can refine the second idea to reduce the size of $G$.
Let $X=\{(l_1,r_1),\dots, (l_t,r_t)\}$ be a feasible solution for the original string.
Let $T_X$ be a tree on the vertex set $\{(l_1,r_1),\dots, (l_t,r_t)\}\cup \{(0,n+1)\}$ such that $(l,r)$ is an ancestor of $(l',r')$ if and only if $[l',r']\subseteq [l,r]$.
We then make use of the heavy-light decomposition of $T_X$ in our construction.

Next, we consider the following construction.
As with the second idea, a vertex in $G$ corresponds to a list of pairs $((l_1,r_1),\ldots,(l_k,r_k))$ with $(A_{l_i},A_{r_i})\in \mathcal{R}$ for every $i$ such that $l_i < l_j < r_j < l_i$ for every $i < j$, but we do not introduce vertices for all such lists.
Instead, we only keep the ones in which the pairs in the list represent the light edges along the path from the root to $(l_k,r_k)$ in a tree  $T_X$ for some solution $X$.
The important observation here is that because there are at most $\log n$ light edges on a path in $T$, $|V|$ is bounded by $n^{O(\log n)}$. Therefore, by applying Theorem~\ref{thm:main} on $G$, we can obtain a polylogarithmic average sensitivity bound.


\begin{algorithm}[t!]
\caption{Algorithm for RNA Folding Problem Modeled by the MWC Problem}\label{alg:RNA_ours}
\Procedure{\emph{\Call{RecRNA}{$I$}}}{
    \For{$H=[l(H),r(H)]\subseteq [l(I)+1,r(I)-1]$ s.t.\ $A_{l(H)}=A_{r(H)}$}{
        $\Tmp[H][l(I)]\leftarrow \{(l(H),r(H))\}$\;
        \For{$r=l(I)+1,\dots, l(H)-1$}{
            $(i^*,L^*)\leftarrow \mathrm{argmax}_{\substack{(i,L)\colon L=[l(L),r]\subseteq [l(I)+1,r(I)+1]\setminus H, l(I)<i<l(L),\\ \left(A_{l(L)},A_{r}\right)\in \mathcal{R}, r-l(L)\leq r(H)-l(H)}}\Bigl|\Call{RecRNA}{L}\Bigr|$\;
            $\Tmp[H][r]\leftarrow \Tmp[H][i^*]\cup \{(l(L^*),r(L^*))\}\cup \Call{RecRNA}{L^*}$\label{line:trans_1}\; \Comment{Reuse $\Call{RecRNA}{L^*}$ computed above.}
        }
        \For{$r=r(H)+1,\dots, l(I)-1$}{
            $(i^*,L^*)\leftarrow \mathrm{argmax}_{\substack{(i,L)\colon L=[l(L),r]\subseteq [l(I)+1,r(I)-1]\setminus H, I(I)\leq i<l(L), r(H)<l(L),\\ \left(A_{l(L)},A_{r}\right)\in \mathcal{R}, r-l(L)\leq r(H)-l(H)}}\Bigl|\Call{RecRNA}{L}\Bigr|$\;
            $\Tmp[H][r]\leftarrow \Tmp[H][i^*]\cup \{(l(L^*),r(L^*))\}\cup \Call{RecRNA}{L^*}$\;\label{line:trans_2} \Comment{Reuse $\Call{RecRNA}{L^*}$ computed above.}
        }
        $R(H)\leftarrow \Call{RecRNA}{H}\cup \Tmp\left[\mathrm{argmax}_{i\in [l(I),r(I)-1]}\Bigl|\Tmp[H][i]\Bigr|\right]$
    }
    $H^{*}\leftarrow \mathrm{argmax}_{\substack{H=[l(H),r(H)]\subseteq [l(I)+1,r(I)-1],\\ \left(A_{l(H)},A_{r(H)}\right)\in \mathcal{R}}}R(H)$\;\label{line:argmax2}
    \Return $R(H^{*})$\;
}
\Procedure{\emph{\Call{MWCRNA}{$A$}}}{
    \Return $\Call{RecRNA}{[0,n+1]}$\;
}
\end{algorithm}

The graph we construct to apply Algorithm~\ref{alg:main} on is designed to represent the computation of the DP algorithm given in $\Call{MWCRNA}{A}$ of Algorithm~\ref{alg:RNA_ours}.
For an interval $I$, let $l(I)$ and $r(I)$ denote the left and right ends of $I$, respectively.
For an interval $I$, $\Call{RecRNA}{I}$ computes an optimal solution that matches bases in $[l(I)+1,r(I)-1]$. 
A call of $\Call{RecRNA}{I}$ first tries all possibility of an interval $H$, which is a heavy child of $I$ in tree $T$. Then, we compute the optimal way to match bases in $[l(I)+1,r(I)-1]\setminus H$ via DP using a recursive call of $\Call{RecRNA}{L}$, where $L$ is an interval corresponding to a light child of $I$. Finally, we recursively call $\Call{RecRNA}{H}$ on the heavy child $H$ and take a solution of maximum weight over all possibility of $H$.

In Algorithm~\ref{alg:RNA_ours}, $\Tmp[l(H),r(H)][i]$ in $\Call{RecRNA}{I}$ represents a solution with maximum possible weight under the constraints
\begin{itemize}
    \item $(l(I),r(I))$ is matched, 
    \item $(l(H),r(H))$ is matched, 
    \item the way to match the bases in $[l(I)+1,i]\setminus [l(H),r(H)]$ is already determined, and
    \item all remaining bases, which are the bases in $[0,n+1]\setminus I$ and $[i+1,l(I)-1]\cup [l(H)+1,r(H)-1]\setminus \{l(H),r(H)\}$ are not matched.
\end{itemize}
The transition in Line~\ref{line:trans_1} and Line~\ref{line:trans_2} decides to match the bases in $[l(L),r(L)]$. Here, Line~\ref{line:trans_1} considers the cases $r(L)<l(H)$ and Line~\ref{line:trans_2} considers the cases $r(H)<l(L)$.

Let us formally construct the graph $G=(V,E,w)$ so that Algorithm~\ref{alg:main} can be applied on it.
A \emph{pseudo-interval} is either an interval or $\emptyset$. The pseudo-interval $I'$ is \emph{strictly inside} another interval $I$ if $I'\subseteq I$ and $\{l(I),r(I)\}\cap I'=\emptyset$, i.e., $I'$ is contained in $I$ but does not contain the endpoints of $I$.
A triple $(I,H,L)$ of pseudo-intervals is \emph{well-ordered} if all the following conditions hold:
\begin{enumerate}
    \itemsep=0pt
    \item[(a)] $I\neq \emptyset$ and $l(I)<r(I)$.
    \item[(b)] $H\neq \emptyset$ and $l(H)<r(H)$.
    \item[(c)] $H$ and $L$ are strictly inside $I$.
    \item[(d)] $H\cap L=\emptyset$.
\end{enumerate}
Intuitively, $(I,H,L)$ encodes a light edge $(I,L)$ in $T_X$ for some solution $X$, where $I$ is a parent of $L$, and $H$ represents the heavy child of $I$.

Our reduction is randomized and first samples a parameter $B$ from the uniform distribution over $[\log n, 2\log n]$.
Each vertex in $G$ is represented by a list of well-ordered pseudo-interval triples $((I_1,H_1,L_1),\dots,(I_k,H_k,L_k))$ of length at most $B$ that satisfies all the following conditions.
\begin{enumerate}
    \itemsep=0pt
    \item[(i)] For each $j$, $I_j\subseteq [0,n+1]$ and $H_j,L_j\subseteq [1,n]$.
    \item[(ii)] For each $j$, $\left(A_{l(H_j)},A_{r(H_j)}\right)\in \mathcal{R}$ holds. 
    \item[(iii)] For each $j$ with $L_j\neq \emptyset$, $\left(A_{l(L_j)},A_{r(L_j)}\right)\in \mathcal{R}$ holds.
    \item[(iv)] For each $j<j'$, $I_{j'}\subseteq L_j$ hold.
\end{enumerate}
Intuitively, the vertex $((I_1,H_1,L_1),\dots,(I_k,H_k,L_k))$ represents a path in $T_X$ for some solution $X$ from the root to $L_k$ if $L_k\neq\emptyset$ and to $H_k$ otherwise.
Moreover, for each $j$ with $L_j\neq \emptyset$, $(I_j,H_j,L_j)$ represents the $j$-th light edge $(I_j,L_j)$ on the path from the root in $T_X$.
More specifically, the first three conditions ensure that each $I_j$, $H_j$ and $L_j$ can appear as a vertex of $T_X$ for some $X$. Note that, to consider the dummy vertex $[0,n+1]$ that appears at the root of $T_X$, we allow $I_j\subseteq [0,n+1]$, not $I_j\subseteq [1,n]$. The last condition ensures that, the parent of the $j'$-th light edge is indeed a descendant of the $j$-th light edge for $j<j'$.

For each vertex $v\in V$, the weight $w(v)$ of $v$ is set to $1$.

Let us define the edge set $E$ of $G$.
To make $G$ acyclic and transitive, we need a partial order over $V$.
First, we define a partial order $\preceq$ over well-ordered pseudo-interval triples.
Specifically, for two well-ordered pseudo-interval triples $(I,H,L)$ and $(I',H',L')$, $(I,H,L) \preceq (I',H',L')$ holds if
\begin{itemize}
    \itemsep=0pt
    \item $(I,H,L)=(I',H',L')$, or
    \item $I'\subseteq H$, or
    \item $I=I'$, $H=H'$ and $L=\emptyset$, or
    \item $I=I'$, $H=H'$, $L\neq \emptyset$, $L'\neq \emptyset$ and $r(L)<l(L')$.
\end{itemize} 
Meanwhile, these conditions imply $I'\subseteq I$ because $H\subseteq I$.
Intuitively, $(I,H,L)\preceq (I',H',L')$ holds if we traverse the light edge $(I,L)$ before the light edge $(I',L')$ in a fixed pre-order transversal of $T_X$ for some solution $X$.
The pre-order transversal of a tree depends on the order of the children of vertices. 
Here, we first traverse the light children $L$ in ascending order of $l(L)$, and we then we traverse the heavy child.

We introduce an edge from the vertex $((I_1,H_1,L_1),\dots,(I_k,H_k,L_k))$ to $((I'_1,H'_1,L'_1),\dots,(I'_{k'},H'_{k'},L'_{k'}))$ if the former is lexicographically strictly smaller than the latter, where we compare triples by the order $\preceq$.
The next lemma ensures that $\preceq$ is indeed a partial order.
This also ensures that $G$ is acyclic and transitive.
\begin{lemma}
$\preceq$ is a partial order.
\end{lemma}
\begin{proof}
The reflexivity of $\preceq$ is clear from the definition.
Now, we prove $\preceq$ is antisymmetric.
Assume $(I,H,L)\neq (I',H',L')$ satisfies $(I,H,L)\preceq (I',H',L')\preceq (I,H,L)$. Then, we have $I\subseteq I'\subseteq I$ and therefore $I=I'$ and $H=H'$. If $L=\emptyset$, we have $L'=\emptyset$ because of $(I',H',L')\preceq (I,H,L)$ and therefore we have $(I,H,L)=(I',H',L')$. Finally, if $L\neq \emptyset$ and $L'\neq \emptyset$, we have $r(L)<l(L')<r(L')<l(L)<r(L)$, which is a contradiction.
Therefore $\preceq$ is asymmetric.

Finally, we prove $\preceq$ is transitive.
Assume $(I,H,L)\neq (I',H',L')\neq (I'',H'',L'')$ satisfies $(I,H,L)\preceq(I',H',L')\preceq(I'',H'',L'')$.
We prove that $(I,H,L)\preceq (I'',H'',L'')$.
If $I'\subseteq H$, we have $I''\subseteq I'\subseteq H$.
Therefore we have $I''\subseteq H$ and the claim holds.
Otherwise, we have $I=I'$ and $H=H'$. If $I''\subseteq H'$, we have $I''\subseteq H'=H$. Therefore we have $I''\subseteq H$ and the claim holds.
Now, we can assume $I=I'=I''$ and $H=H'=H''$. If $L=\emptyset$, the claim holds. Otherwise, none of $L,L',L''$ is empty and therefore $r(L)<l(L')<r(L')<l(L'')$. Then, we have $r(L)<l(L'')$ and the claim holds.
Therefore, $\preceq$ is transitive and thus is thus a partial order.
\end{proof}

The entire algorithm is given in Algorithm~\ref{alg:RNA}.

\begin{algorithm}[t!]
\caption{Stable-on-average RNA folding}\label{alg:RNA}
\Procedure{\emph{\Call{ConstructGraph}{A}}}{
    Sample $B$ from the uniform distribution from $[\log n, 2\log n]$\;
    Let $V(G)$ be the set of all lists of pseudo-interval triples of length at most $B$ that satisfies all conditions~(i), (ii), (iii), and (iv)\;
    \For{$v,v'\in V(G)$}{
        \If{$v$ is lexicographically smaller than $v'$ when pseudo-interval triples are compared in the order $\prec$}{
            Add an edge $(v,v')$ to $G$\;
        }
    }
    \Return $G$\;
}
\Procedure{\emph{\Call{RNAFolding}{A}}}{
    Let $G=\Call{ConstructGraph}{A}$\;
    Let $X$ be an empty set\;
    \ForEach{$((I_1,H_1,L_1),\dots,(I_k,H_k,L_k))\in \Call{MWC}{G}$}{
        \If{$L_k\neq \emptyset$}{
            Add $(l(L_k),r(L_k))$ to $X$\;
        }
        \Else{
            Add $(l(H_k),r(H_k))$ to $X$\;
        }
    }
    \Return $X$\;
}
\end{algorithm}

\subsection{Mapping from Chains to Solutions}

Let us establish a surjective map from chains in $G$ to solutions for the original problem.
Let $P=(v_1,\dots, v_t)$ be a chain in $G$, and let $v_i=((I_{i,1},H_{i,1},L_{i,1}),\dots,(I_{i,k_i},H_{i,k_i},L_{i,k_i}))$ for each $i=1,\ldots,t$.
The solution to which $P$ is mapped is obtained such that for each $i$, matching two endpoints of $H_{i,k_i}$ if $L_{i,k_i}=\emptyset$ and those of $L_{i,k_i}$ otherwise.
The next lemma ensures that the solution obtained this way does not contain a pseudoknot and two identical pairs of endpoints. 
\begin{lemma}
Let $v=((I_1,H_1,L_1),\dots, (I_k,H_k,L_k))$, $v'=((I'_1,H'_1,L'_1),\dots, (I'_{k'},H'_{k'},L'_{k'}))$ and suppose that $v$ is lexicographically strictly smaller than $v'$.
Then, each of the pseudo-interval pairs $(H_k,H'_{k'})$, $(H_k,L'_{k'})$, $(L_k,H'_{k'})$, and $(L_k,L'_{k'})$ satisfies one of the following: one of the two pseudo-intervals is strictly inside the other, they are disjoint, or they coincide.
Moreover, the third case happens only when $I_k=I'_{k'}$ and $L'_{k'}\neq \emptyset$ holds, in which $H_k = H'_{k'}$.
\end{lemma}
\begin{proof}
If $v$ is a prefix of $v'$, we have
\[
    I'_{k'}\subseteq L'_{k}=L_{k},
\]
where the set inequality is from $k<k'$ and condition~(iv) and the equality is from the assumption that $v$ is a prefix of $v'$. Therefore, we conclude that both of $H'_{k'}, L'_{k'}$ are strictly inside $L_{k}$ and $H_{k}\cap H'_{k'}=H_{k}\cap L'_{k'}=\emptyset$.

Assume $v$ is not a prefix of $v'$ and let $j$ be the first index such that $(I_{j},H_{j},L_{j})\prec (I'_{j},H'_{j},L'_{j})$.
From the definition of $\prec$, we have $I'_{j}\subseteq H_{j}$ or $(I_{j},H_{j})=(I'_{j},H'_{j})$. If $I'_{j}\subseteq H_{j}$ and $j=k$, we have
\[
    I'_{k'}\subseteq I'_{j}\subseteq H_{j} = H_{k},
\]
where the first set inequality is from $j\leq k'$ and the equality is from $j=k$.
Therefore, both $H'_{k'}, L'_{k'}$ are strictly inside $H_{k}$ and $L_{k}\cap H'_{k'}=L_{k}\cap L'_{k'}=\emptyset$.
If $I'_{j}\subseteq H_{j}$ and $j<k$, we have
\[
    I_{k}\cap I'_{k'}
    \subseteq I_{k}\cap I'_{j}
    \subseteq I_{k}\cap H_{j}
    \subseteq H_{j}\cap L_{j}=\emptyset,
\]
where the first set inequality is from $j\leq k'$, the second set inequality is from $I'_{j}\subseteq H_{j}$, the third set inequality is from $j<k$ and condition~(iv), and the equality is from the condition (d).
Therefore we have $H_{k}\cap H'_{k'}=H_{k}\cap L'_{k'}=L_{k}\cap H'_{k'}=L_{k}\cap L'_{k'}=\emptyset$.

Assume $(I_{j},H_{j})=(I'_{j},H'_{j})$. If $j<k$, we have $I_k\subseteq L_j$ from condition (iv). Therefore, $H_k$ and $L_k$ are contained in $L_j$, except for the case $j=k$, which $H_k=H_j$ holds. Similarly, $H'_{k'}$ and $L'_{k'}$ are contained in $L'_j$, except for the case $j=k'$, which $H_{k'}=H_j$ holds.
Now, we have $H_j\cap L'_j=H'_j\cap L'_j=\emptyset$, $L_j\cap H'_j=L_j\cap H_j=\emptyset$ and $L_j\cap L'_j=\emptyset$, where the last claim emrges from the fact that $r(L_j)<l(L'_j)$ holds unless $L_j=\emptyset$. Therefore, we have $H_{k}\cap H'_{k'}=H_{k}\cap L'_{k'}=L_{k}\cap H'_{k'}=L_{k}\cap L'_{k'}=\emptyset$ for almost all cases. The only exception is that $H_{k}=H_{k'}$ holds if $j=k=k'$ and in this case, we have $L'_{k'}\neq \emptyset$ because of $(I_{j},H_{j},L_{j})\prec (I'_{j},H'_{j},L'_{j})$. 
Thus, the lemma is proved.
\end{proof}

Therefore, the solution we obtained from $P$ is feasible. 
Moreover, this solution preserves the weight of the chain $P$.

Next, we prove that this map is indeed surjective.
Let $X=\{(l_1,r_1),\dots, (l_t,r_t)\}$ be a solution for the original problem. We construct a chain in $G$ that is mapped to $X$.
Let $T_X$ be a tree such that the vertex set is $\{(l_1,r_1),\dots, (l_t,r_t)\}\cup \{(0,n+1)\}$ and $(l,r)$ is an ancestor of $(l',r')$ if and only if $[l',r']\subseteq [l,r]$. 
Let us fix a heavy-light decomposition of $T_X$.
For a tree $T_X$ and its heavy-light decomposition, the desired chain in $G$ can be obtained by using $\Call{MakePath}{T_X}$ given in Algorithm~\ref{alg:makepath}.

\begin{algorithm}[t!]
\caption{Chain construction}\label{alg:makepath}
\Procedure{\emph{\Call{DFS}{$I$}}}{
    \If{$I$ has no child}{
        \Return\;
    }
    Let $H$ be the heavy child of $I$\;
    Append $(I,H,\emptyset)$ to the end of $\mathsf{CurrentList}$\;
    Append $\mathsf{CurrentList}$ to the end of $\mathsf{Chain}$\;\label{line:appendheavy}
    Remove $(I,H,\emptyset)$ from the end of $\mathsf{CurrentList}$\;
    \ForEach{light children $L$ of $I$ in increasing order of $l(L)$}{
        Append $(I,H,L)$ to the end of $\mathsf{CurrentList}$\;
        Append $\mathsf{CurrentList}$ to the end of $\mathsf{Chain}$\;\label{line:appendlight}
        \Call{DFS}{$L$}\;
        Remove $(I,H,L)$ from the end of $\mathsf{CurrentList}$\;
    }
    \Call{DFS}{$H$}\;
}
\Procedure{\emph{\Call{MakeChain}{$T_X$}}}{
    Let $\mathsf{CurrentList}$ and $\mathsf{Chain}$ be an empty list\;
    \Call{DFS}{$[0,n+1]$}\;
    \Return $\mathsf{Chain}$\;
}
\end{algorithm}

We verify $\Call{MakeChain}{T_X}$ in Algorithm~\ref{alg:makepath}  outputs a chain in $G$.
First, we prove that each list in $\Call{MakeChain}{T_X}$ is a vertex of $G$.
\begin{lemma}
Each list $((I_1,H_1,L_1),\dots, (I_k,H_k,L_k))$ in $\Call{MakeChain}{T_X}$ consists of only well-ordered triples, satisfies the conditions (i), (ii), (iii) and (iv), and has length at most $\log n$.
\end{lemma}
\begin{proof}
For $j=1,\dots, k$, it is clear from the algorithm that $(I_j,H_j,L_j)$ satisfies all the conditions (a), (b), (c) and (d). Therefore, each triple is well-ordered. The conditions (i), (ii), (iii), (iv) are also clear from the algorithm. The remaining problem is to bound the length $k$ of the list.

Throughout the algorithm, the length of $\mathsf{CurrentList}$ increases only when we call \Call{Rec}{$L$}.
From the definition of a heavy child, the size of the subtree rooted at $L$ is less than half of that of $I$.
Since the number of the vertices of $T$ is at most $\frac{n}{2}$, the length of $\mathsf{CurrentList}$ can be at most $\log n-1$ and hence we have $k\leq \log n$. Therefore, the lemma is proved.
\end{proof}

Now we prove that $\Call{MakeChain}{T_X}$ outputs a chain.
\begin{lemma}
$\Call{MakeChain}{T_X}$ outputs a chain in $G$.
\end{lemma}
\begin{proof}
Let $v=((I_1,H_1,L_1),\dots, (I_k,H_k,L_k))$ and $v'=((I'_1,H'_1,L'_1),\dots, (I'_{k'},H'_{k'},L'_{k'}))$ in $\Call{MakeChain}{T_X}$ such that $v'$ comes next from $v$.
It suffices to show that $v'$ is lexicographically larger than $v$ when we compare pseudo-intervals by order $\prec$.
If $v$ is a prefix of $v'$, then the claim is clear.
Furthermore, $v'$ cannot be a prefix of $v$ because if it were, $v$ and $v'$ should be added to $\mathsf{Chain}$ during and before executing $\Call{DFS}{I_{k'}}$, respectively.
Now, we assume $v$ is not a prefix of $v'$, and we let $j$ be the first index such that $(I_j,H_j,L_j)\neq (I'_j,H'_j,L'_j)$.

Let $I^*$ be the shortest interval such that $I_k\subseteq I^*$, $I'_{k'}\subseteq I^*$ and $\Call{DFS}{I^*}$ is called. Then, both $(I_k,H_k,L_k)$ and $(I'_{k'},H'_{k'},L'_{k'})$ are appended to $\mathsf{CurrentList}$ during the execution of $\Call{DFS}{I^*}$. Since $\Call{DFS}{I^*}$ appends at least two lists to $\mathsf{Chain}$, $I^*$ has at least one child.
Let $H^*$ be the heavy child of $I^*$ and let $L^*_1,\dots,L^*_s$ be the light children of $I^*$, where $r(L^*_p)<l(L^*_{p+1})$ holds for all $p=1,\dots, s-1$.

Let us closely look at how $\Call{DFS}{I^*}$ works. Observe that $v'$ is not the first list that is appended to $\mathsf{Chain}$ during the execution of $\Call{DFS}{I^*}$. Thus, $v'$ is appended to $\mathsf{Chain}$ either in Line~\ref{line:appendlight} in $\Call{DFS}{I^*}$ or in Line~\ref{line:appendheavy} of $\Call{DFS}{H^*}$.

Assume the former. Then, we have $k'=j$. Let $(I'_{k'},H'_{k'},L'_{k'})=(I'_j,H'_j,L'_j)=(I^*,H^*,L^*_p)$. If $p=1$, we have $(I_j,H_j,L_j)=(I^*,H^*,\emptyset)\prec (I^*,H^*,L^*_1)=(I'_j,H'_j,L'_j)$ holds. Otherwise, $(I_j,H_j,L_j)=(I^*,H^*,L^*_{p-1})\prec (I^*,H^*,L^*_p)=(I'_j,H'_j,L'_j)$ holds. Therefore we have $(I_j,H_j,L_j)\prec (I'_j,H'_j,L'_j)$ and $v\prec v'$.

Assume the latter. Then, we have $I'_j=H^*=H_j$. Therefore we have $(I_j,H_j,L_j)\prec (I'_j,H'_j,L'_j)$ and $v\prec v'$.
\end{proof}


\subsection{Pseudo-antichain}

Now, we consider deleting a letter from the original string and define $S_i$'s so that the third condition listed at the beginning of Section~\ref{sec:application} is satisfied.
For $i=1,\dots, n$, let $S_i$ be the set of vertices $v$ in the graph $G=(V,E,w)$ constructed such that the index $i$ is ``relevant'' to $v$.
Formally, the vertex $((I_1,H_1,L_1),\dots,(I_k,H_k,L_k))$ is in $S_i$ if $i$ is an endpoint of at least one of $I_1,H_1,L_1,\dots, I_k,H_k,L_k$.
Since the edge set $E$ is defined only by the inequality relation over integers, the graph induced by $V\setminus S_i$ is isomorphic to the graph constructed from the string $A^i:=A_1\dots A_{i-1}A_{i+1}\dots A_n$.

If $S_i$ were an antichain, we would apply Theorem~\ref{thm:main} to obtain a stable-on-average algorithm for RNA folding problem.
Unfortunately, $S_i$ is not an antichain in general.
However, $S_i$ has a property similar to an antichain.
We will prove that, for any $v\in V(G)$, $S_i$ crosses both of $V_{-v}(G)$ and $V_{+v}(G)$ only if $v\in S_i$. 
To prove this, we characterize indices $i$ with $V_{-v}(G)\cap S_i\neq \emptyset$ by the following lemma:
\begin{lemma}\label{lem:RNA_before}
Let $i\in \{1,\dots, n\}$ and $v=((I_1,H_1,L_1),\dots,(I_k,H_k,L_k))\in V(G)\setminus S_i$.
Suppose that there is a vertex $v'=((I'_1,H'_1,L'_1),\dots,(I'_{k'},H'_{k'},L'_{k'}))\in S_i$ with $v'\prec v$.
Then, $i$ satisfies (exactly) one of the following conditions:
\begin{itemize}
    \item $i\not \in I_1$,
    \item $i\in L_{j-1}\setminus I_j$ holds for some $j$, or
    \item $i\in I_j\setminus (H_j\cup L_j)$, $L_j\neq \emptyset$ and $i<l(L_j)$ holds for some $j$.
\end{itemize}
\end{lemma}
\begin{proof}
Let $j'$ be an index such that $i$ appears as an endpoint of one of $I'_{j'}$, $H'_{j'}$ or $L'_{j'}$.
Then, $v'$ is not a prefix of $v$ because $(I_{j'},H_{j'},L_{j'})$ cannot appear in a vertex in $S_i$. Let $j\leq j'$ be the first index with $(I'_j,H'_j,L'_j)\prec (I_j,H_j,L_j)$.
From the definition of $\prec$, we have $I_j\subseteq H'_j$ or $(I_j,H_j)=(I'_j,H'_j)$.

Assume $I_j\subseteq H'_j$.
If $j<j'$, then we have $i\in I'_{j'}\subseteq L'_j\subseteq I'_j\setminus H'_j\subseteq I'_j\setminus I_j$,
where the first set inequality is from $j<j'$, the second set inequality is from well-orderedness, and the last set inequality is from $I_j\subseteq H'_j$. If $j=j'$, then $i$ is an endpoint of either $I'_j$, $H'_j$ or $L'_j$.
Thus, $i\in I'_j\setminus I_j$, because $I_j$ is contained in $H'_j$ and cannot have $i$ as an endpoint.
Therefore, in both cases, we have $i\in I'_j\setminus I_j\subseteq L_{j-1}\subseteq I_j$ for $j>1$ and $i\not \in I_j$ for $j=1$.

Now, assume $(I_j,H_j)=(I'_j,H'_j)$. In this case, we have $L_j\neq \emptyset$ by definition of $\prec$. 
Since $i$ can neither be an endpoint of $I'_j$ nor $H'_j$, we have $i\in L'_j$.
Therefore, from the definition of $\prec$, we have $i\in L'_j\subseteq I_j\setminus (H_j\cup L_j)$ and $i\leq r(L'_{j})<l(L_{j})$ and the lemma is proved.
\end{proof}

Next, we characterize indices $i$ with $V_{+v}(G)\cap S_i\neq \emptyset$.
\begin{lemma}\label{lem:RNA_after}
Let $i\in \{1,\dots, n\}$ and $v=((I_1,H_1,L_1),\dots,(I_k,H_k,L_k))\in V(G)\setminus S_i$.
Suppose that there is a vertex $v'=((I'_1,H'_1,L'_1),\dots,(I'_{k'},H'_{k'},L'_{k'}))\in S_i$ with $v\prec v'$. Then, $i$ satisfies one of the following conditions:
\begin{itemize}
    \item $i\in L_k$,
    \item $i\in H_j$ holds for some $j$,
    \item $i\in I_j\setminus (H_j\cup L_j)$ and $L_j = \emptyset$ holds for some $j$, or
    \item $i\in I_j\setminus (H_j\cup L_j)$, $L_j\neq \emptyset$ and $r(L_j)<i$ holds for some $j$.
\end{itemize}
\end{lemma}
\begin{proof}
Let $j'$ be an index that $i$ appears as an endpoint of one of $I'_{j'}$, $H'_{j'}$ or $L'_{j'}$. If $v$ is a prefix of $v'$, we have $k<j'$ because $(I'_{j'},H'_{j'},L'_{j'})$ cannot appear as one of the triples in $v$.
Therefore, we have $i\in I'_{j'}\subseteq L'_k=L_k$, where the set inequality is from the condition (iv) and the equality is from the assumption that $v$ is a prefix of $v'$.

Now we assume that $v$ is not a prefix of $v'$, and we let $j\leq j'$ be the first index with $(I_j,H_j,L_j)\prec (I'_j,H'_j,L'_j)$. From the definition of $\prec$, we have $I'_j\subseteq H_j$ or $(I_j,H_j)=(I'_j,H'_j)$.

If $I'_j\subseteq H_j$, we have $i\in I'_{j'}\subseteq I'_j\subseteq H_j$. Now, we assume $(I_j,H_j)=(I'_j,H'_j)$. Since $i$ can neither be an endpoint of $I'_j$ nor $H'_j$, we have $i\in L'_j\subseteq I_j\setminus (H_j\cap H'_j)$. Furthermore, if $L_j\neq \emptyset$, we have $r(L_j)<l(L'_j)\leq i$ from the definition of $\prec$. Therefore, the lemma is proved.
\end{proof}

We can observe that the conditions in Lemma~\ref{lem:RNA_before} and Lemma~\ref{lem:RNA_after} are disjoint. Indeed, if $i\in I_1$, we can take the last index $j$ with $i\in I_j$. Then, exactly one of $i\in H_j$, $i\in L_j$ or $i\in I_j\setminus (H_j\cup L_j)$ holds. If $i\in H_j$, there is nothing to state. If $i\in L_j$, unless $j=k$, we have $i\in L_j\setminus I_{j+1}$ because $j$ is the last index wherein $I_j$ contains $i$. Finally, If $i\in I_j\setminus (H_j\cup L_j)$, we have either $L_j=\emptyset$, $i<l(L_j)$ or $r(L_j)<i$ because $i$ is not in $L_j$.
Therefore we have the following.
\begin{lemma}\label{lem:RNA_beforeafter}
Let $i\in \{1,\dots, n\}$ and $v\in V(G)$. Then, at least one of $V_{-v}(G)\cap S_i=\emptyset$, $V_{+v}(G)\cap S_i=\emptyset$ or $v\in S_i$ holds. Moreover, for any $v\in V(G)$, there are at most $6B\leq 12\log n$ indices $i$ that satisfy both $V_{-v}(G)\cap S_i\neq \emptyset$ and $V_{+v}(G)\cap S_i\neq \emptyset$.
\end{lemma}

\subsection{Proof of Theorem~\ref{thm:rna-folding}}

We extend the analysis of Algorithm~\ref{alg:main} to handle the case in which potentially missing sets $S_i$ are not necessarily antichains, but they satisfy Lemma~\ref{lem:RNA_beforeafter}. 
The discussion in Section~\ref{sec:pivot} does not depend on the fact that potentially missing sets are antichains.
Thus, the same proof as in Section~\ref{sec:pivot} goes through, and the claim of Lemma~\ref{lem:pivottvd} holds even when $S_i$ are not antichains.
Thus as in Section~\ref{sec:proofstrategy}, we focus on proving the claim of Lemma~\ref{lem:main} when $S_i$'s satisfy Lemma~\ref{lem:RNA_beforeafter}.

Let us fix the random bits used in \Call{Rec}{$V,\epsilon$} in Algorithm~\ref{alg:main}. Let $\mathcal{U}_j$ be the family of all sets $U$ such that \Call{Rec}{$U,\epsilon$} is called in a recursion step of depth $j$ for $j=0,\dots, k$, where $k$ is the maximum index $j$ with $\mathcal{U}_j\neq \emptyset$.
By the same observation as described in Section~\ref{sec:proofstrategy}, we have $k = O(\log |V|)$.

For a set $U\in \mathcal{U}_j$, let $n_U$ be the number of the potentially missing sets with $S_i\cap U\neq \emptyset$. 
We bound the sum of $n_U$ over $j\in \{0,\dots, k\}$ and $U\in \mathcal{U}_j$.
First, we prove the following.
\begin{lemma}\label{lem:RNA_partsum}
Let $j\in \{0,\dots, k\}$ and $\mathcal{W}$ be a subfamily of $\mathcal{U}_j$. Then, we have
\[
    \sum_{U\in \mathcal{W}}n_U\leq n+|\mathcal{W}|\cdot 12\log n.
\]
\end{lemma}
\begin{proof}
We prove by induction on $j$. If $j=0$, the claim is clear. Assume $j\geq 1$. For each set $U\in \mathcal{W}$, let the \emph{parent} of $U$ be the unique set $U'\in \mathcal{U}_{j-1}$ with $U\subseteq U'$. Let $\mathcal{W}'$ be the family of sets that is a parent of some set in $\mathcal{W}$.
From the construction, each set in $\mathcal{W}'$ is the parent of one or two sets in $\mathcal{W}$.
If $U'\in \mathcal{W}'$ is the parent of exactly one set $U\in \mathcal{W}$, we have $n_{U}\leq n_{U'}$.
Similarly, if $U'\in \mathcal{W}'$ is the parent of exactly two sets $U_1,U_2\in \mathcal{W}$, we have $n_{U_1}+n_{U_2}\leq n_{U'}+12\log n$.
Therefore, we have
\begin{align*}
    \sum_{U\in \mathcal{W}}n_U
    &\leq \sum_{U'\in \mathcal{W}}n_{U'}+\left(|\mathcal{W}|-|\mathcal{W}'|\right)\cdot 12\log n\\
    &\leq n+|\mathcal{W}'|\cdot 12\log n+\left(|\mathcal{W}|-|\mathcal{W}'|\right)\cdot 12\log n=n+|\mathcal{W}|\cdot 12\log n,
\end{align*}
where the first inequality is from the above observation and the second inequality is from the induction hypothesis.
\end{proof}

Now, we have the following bound on the sum of $n_U$:
\begin{lemma}\label{lem:RNA_recnsum}
We have
\[
    \sum_{j=0}^{k}\sum_{U\in \mathcal{U}_j}n_U = O\left(n\log |V(G)|\right). 
\]
\end{lemma}
\begin{proof}
First, we have $\sum_{j=0}^{k}|\mathcal{U}_j|\leq \frac{n}{2}$, since any chain in $G$ contains at most $\frac{n}{2}$ vertices and each call of \Call{Rec}{$U$} add exactly one vertex to the output.
Therefore, we have
\begin{align*}
    \sum_{j=0}^{k}\sum_{U\in \mathcal{U}_j}n_U
    \leq \sum_{j=0}^{k}\left(n+|\mathcal{U}_j|\cdot 12\log n\right)
    \leq (k+1)n+\frac{n}{2}\cdot 12\log n = O(n\log |V(G)|),
\end{align*}
where the first inequality is from Lemma~\ref{lem:RNA_partsum}, the second inequality is from $\sum_{j=0}^{k}|\mathcal{U}_j|\leq \frac{n}{2}$ and the last inequality is from $k\leq \log |V(G)|$ and $n\leq |V(G)|$.
\end{proof}

Now for fixed $\epsilon$, the average sensitivity of $\Call{Rec}{}$ is bounded by
\begin{align}
    \frac{1}{n}\sum_{i=1}^{n}\sum_{j=0}^{k_i}\sum_{U\in \mathcal{U}_j, U\cap S_i\neq \emptyset}\TV\left(\bar{v}(U),\bar{v}(U\setminus S_i)\right)\cdot n_{U},\label{eq:RNA_recemd}
\end{align}
where $\bar{v}(U)$ is the random variable of the pivot chosen in $\Call{Rec}{U}$. 
Then, we have the following:
\begin{lemma}\label{lem:RNA_main}
We have
\begin{align*}
    &\frac{1}{n}\sum_{i=1}^{n}\EM\left(\Call{MWC}{G}, \Call{MWC}{G[V\setminus S_i]}\right)
    = O\left(K\epsilon^{-1}\log |V|\log\left(|V|\epsilon^{-1}\right)\right),
\end{align*}
where $K=12\log n$.
\end{lemma}
\begin{proof}
We have
\begin{align*}
    &\frac{1}{n}\sum_{i=1}^{n}\EM\left(\Call{MWC}{G}, \Call{MWC}{G[V\setminus S_i]}\right)\\
    &\leq \frac{1}{n}\sum_{i=1}^{n}\E\left[\sum_{j=0}^{k}\sum_{U\in \mathcal{U}_j, U\cap S_i\neq \emptyset}\TV\left(\bar{v}(U),\bar{v}(U\setminus S_i)\right)\cdot n_U\right]\\
    &= \sum_{j=0}^{k_i}\E\left[\frac{1}{n}\sum_{U\in \mathcal{U}_j}\left(\sum_{i\colon U\cap S_i\neq \emptyset}\TV\left(\bar{v}(U),\bar{v}(U\setminus S_i)\right)\cdot n_U\right)\right]\\
    &\leq \sum_{j=0}^{k_i}\E\left[\frac{1}{n}\sum_{U\in \mathcal{U}_j}O\left(K\epsilon^{-1}\log\left(|U|\epsilon^{-1}\right)\right)\cdot n_U\right]\\
    &\leq O\left(K\epsilon^{-1}\log |V|\log\left(|V|\epsilon^{-1}\right)\right),
\end{align*}
where the first inequality is from (\ref{eq:RNA_recemd}), the second inequality is from Lemma~\ref{lem:pivottvd}, and the last inequality is from Lemma~\ref{lem:RNA_recnsum}.
\end{proof}

Therefore, for a fixed upper bound $B$ on the length of the list that defines vertices,  we obtain a $(1-\delta)$-approximation algorithm with average sensitivity $O(K\delta^{-1}\log^3 |V|)$ by applying the procedure \Call{MWC}{$G$}.
Finally, we remove the conditioning of $B$.

\begin{lemma}
The procedure $\Call{RNAFolding}{}$ in Algorithm~\ref{alg:RNA} has an average sensitivity $O(K\delta^{-1}\log^3 |V|)$.
\end{lemma}
\begin{proof}
Recall that $A^i$ denotes the substring $A_1\cdots A_{i-1}A_{i+1}\cdots A_n$.
From Lemma~\ref{lem:benri}, we have
\begin{align*}
    &\frac{1}{n}\sum_{i=1}^{n}\EM\left(\Call{RNAFolding}{A}, \Call{RNAFolding}{A^i}\right)\\
    &\leq \frac{1}{\log n}\int_{\log n}^{2\log n}\left(\frac{1}{n}\sum_{i=1}^{n}\EM\left(\left(\Call{MWC}{G}\mid B=b\right), \left(\Call{MWC}{G[V\setminus S_i]}\mid B=b\right)\right)\right)\mathrm{d}b\\
    &\quad \quad +2\cdot \sum_{i=1}^{n}\left|1-\frac{\log (n-1)}{\log n}\right|\\
    &\leq O(K\delta^{-1}\log^3 |V|)+2\cdot \sum_{i=1}^{n}\left|1-\frac{\log (n-1)}{\log n}\right|\\
    &\leq O(K\delta^{-1}\log^3 |V|)+O(1)=O(K\delta^{-1}\log^3|V|),
\end{align*} 
where the first inequality is from Lemma~\ref{lem:benri}, the second inequality is from Theorem~\ref{thm:main}, and the last inequality is from $\frac{\log (n-1)}{\log n}\geq \frac{n-1}{n}$ for $n\geq 2$.
\end{proof}

Theorem~\ref{thm:rna-folding} follows because $\log |V|\leq \log\left(n^{O(\log n)}\right)=O(\log^2 n)$.

\bibliography{bib}

\newpage
\appendix


\section{Proof of Lemma~\ref{lem:benri}}\label{app:app}


For $i=1,\dots, n$, we have
\begin{align}
    &\D(\Call{Alg}{U},\Call{Alg}{U^i}) \nonumber \\
    &\leq \int_{[B,(1+t)B]\cap [B^i,(1+t)B^i]}\biggl(\min\left(\frac{1}{tB},\frac{1}{tB^i}\right)\D\left((\Call{Alg}{U}\mid p=\hat{p}),(\Call{Alg}{U^i}\mid p^i=\hat{p})\right) \nonumber \\
    & \qquad \qquad \qquad \qquad \qquad \qquad+ \max\left(0,\frac{1}{tB}-\frac{1}{tB^i}\right)M\biggr)\mathrm{d}\hat{p} \nonumber\\
    &\quad \quad + \int_{[B,(1+t)B]\setminus [B^i,(1+t)B^i]}\frac{1}{tB}M\mathrm{d}\hat{p} \nonumber \\
    &\leq \int_{B}^{(1+t)B}\frac{1}{tB}\D\left((\Call{Alg}{U}\mid p=\hat{p}),(\Call{Alg}{U^i}\mid p^i=\hat{p})\right)M\mathrm{d}\hat{p} \nonumber \\
    &\quad \quad + \int_{B}^{(1+t)B}\max\left(0,\frac{1}{tB}-\frac{1}{tB^i}\right)M\mathrm{d}\hat{p} + \int_{[B,(1+t)B]\setminus [B^i,(1+t)B^i]}\frac{1}{tB}M\mathrm{d}\hat{p}, \label{eq:uniform2}
\end{align}
where the first inequality is obtained by transporting the probability mass of $\Call{Alg}{U}$ corresponding to the case $p=\hat{p}$ is transported to that of $\Call{Alg}{U^i}$ corresponding to the case $p^i=\hat{p}$.
Now, if $B^i\leq B$, we have
\begin{align*}
    \eqref{eq:uniform2}
    = 0 + \int_{(1+t)B^i}^{(1+t)B}\frac{1}{tB}M\mathrm{d}\hat{p}
    =\frac{(1+t)(B-B^i)}{tB}|U|=\frac{1+t}{t}\cdot \left|1-\frac{B^i}{B}\right|M.
\end{align*}
Otherwise, we have
\begin{align*}
    \eqref{eq:uniform2}
    &= \int_{B}^{(1+t)B}\left(\frac{1}{tB}-\frac{1}{tB^i}\right)M\mathrm{d}\hat{p} + \int_{B}^{B^i}\frac{1}{tB}M\mathrm{d}\hat{p}\\
    &=\left(\left(\frac{1}{tB}-\frac{1}{tB^i}\right)tB+\frac{B^i-B}{tB}\right)M\\
    &=\left(\frac{1}{B^i}+\frac{1}{tB}\right)(B^i-B)n\leq \frac{1+t}{t}\cdot \left|1-\frac{B^i}{B}\right|M.
\end{align*}
Therefore, we have
\begin{align*}
    &\frac{1}{n}\sum_{i=1}^{n}\D(\Call{Alg}{U},\Call{Alg}{U^i})\\
    &\leq \frac{1}{n}\sum_{i=1}^{n}\int_{B}^{(1+t)B}\left(\frac{1}{tB}\D\left((\Call{Alg}{U}\mid p=\hat{p}),(\Call{Alg}{U^i}\mid p^i=\hat{p})\right)\right)\mathrm{d}\hat{p}+\frac{1}{n}\sum_{i=1}^{n}\frac{1+t}{t}\cdot \left|1-\frac{B^i}{B}\right|M\\
    &= \frac{1}{tB}\int_{B}^{(1+t)B}\left(\frac{1}{n}\sum_{i=1}^{n}\D\left((\Call{Alg}{U}\mid p=\hat{p}),(\Call{Alg}{U^i}\mid p^i=\hat{p})\right)\right)\mathrm{d}\hat{p}+\frac{M}{n}\cdot \frac{1+t}{t}\cdot  \sum_{i=1}^{n}\left|1-\frac{B^i}{B}\right|.
    \qed
\end{align*}

\end{document}